\documentclass{aa}  

\usepackage{graphicx}
\usepackage{txfonts}
\usepackage{hyperref}
\usepackage{xcolor}

\usepackage{caption}
\usepackage{subcaption}
\usepackage[export]{adjustbox}

\usepackage{amsmath}
\usepackage{algorithm}
\usepackage[noend]{algpseudocode}
\usepackage{mathrsfs} 

\graphicspath{{./}{figures/}}

\newcommand{\norm}[1]{\left\lVert#1\right\rVert}

\begin{document}

\title{Dynamic and polarimetric VLBI imaging with a multiscalar approach}

\author{H.Müller
        \inst{1}
        \and
        A.P. Lobanov\inst{1}
        }

\institute{Max-Planck-Institut für Radioastronomie,
        Auf dem Hügel 69, Bonn, 53121, Germany\\
        \email{hmueller@mpifr-bonn.mpg.de}, \email{alobanov@mpifr-bonn.mpg.de}
        }
        
\date{Received September 15, 1996; accepted March 16, 1997}

\abstract
{Due to the limited number of antennas and the limited observation time, an array of antennas in Very Long Baseline Interferometry (VLBI) often samples the Fourier domain only very sparsely. Powerful deconvolution algorithms are needed to compute a final image. Recently multiscale imaging approaches such as DoG-HiT were developed to solve the VLBI imaging problem and showed a promising performance: they are fast, accurate, unbiased and automatic.}
{We extend the multiscalar imaging approach to polarimetric imaging, reconstructions of dynamically evolving sources and finally to dynamic polarimetric reconstructions.}
{These extensions (mr-support imaging) utilize a multiscalar approach. The time-averaged Stokes I image is decomposed by a wavelet transform into single subbands. We use the set of statistically significant wavelet coefficients, the multiresolution support, computed by DoG-HiT as a prior in a constrained minimization manner: we fit the single-frame (polarimetric) observables by only varying the coefficients in the multiresolution support.}
{The Event Horizon Telescope (EHT) is a VLBI array imaging supermassive black holes. We demonstrate on synthetic data that mr-support imaging offers ample regularization and is able to recover simple geometric dynamics at the horizon scale in a typical EHT setup. The approach is relatively lightweight, fast and largely automatic and data driven. The ngEHT is a planned extension of the EHT designed to recover movies at the event horizon scales of a supermassive black hole. We benchmark the performance of mr-support imaging for the denser ngEHT configuration demonstrating the major improvements the additional ngEHT antennas will bring to dynamic, polarimetric reconstructions.}
{Current and upcoming instruments offer the observational possibility to do polarimetric imaging of dynamically evolving structural patterns with highest spatial and temporal resolution. State-of-the-art dynamic reconstruction methods can capture this motion with a range of temporal regularizers and priors. With this work, we add an additional, simpler regularizer to the list: constraining the reconstruction to the multiresolution support.} 

\keywords{Techniques: interferometric - Techniques: image processing - Techniques: high angular resolution - Methods: numerical - Galaxies: jets - Galaxies: nuclei}

\maketitle

\section{Introduction}
In Very Long Baseline Interferometry (VLBI) the signals recorded at single antennas are correlated to achieve a spatial resolution that would not be achievable with single-dish instruments. The correlation product of every antenna pair at a fixed time is the Fourier coefficient (visibility) of the true sky brightness distribution with a Fourier frequency determined by the projected spatial vector joining two antennas (baseline). As the Earth rotates during the observing run, baselines rotate on elliptical tracks in the Fourier domain, hence filling up the Fourier plane (uv-plane) continuously. However, due to the limited amount of antennas and the limited amount of observing time, the coverage of Fourier coefficients (uv-coverage) is sparse. VLBI imaging is the problem to recover the true sky brightness distribution from these sparsely covered Fourier coefficients.

It is a long-standing frontline goal in astronomy to recover images of the shadow of a supermassive black hole. The Event Horizon Telescope (EHT) is a globally spanning VLBI array that observes at $230\,\mathrm{GHz}$ (with a recent upgrade to $345\,\mathrm{GHz}$). With the combination of global baselines and short baselines, the EHT achieves the angular resolution that is needed to capture the first image of the black hole shadow in M87 \citep{eht2019a} and in the Milky Way \citep{eht2022a}. The next-generation Event Horizon Telescope (ngEHT) is a planned extension of the EHT \citep{Doeleman2019, Johnson2023}. It may produce movies of the accretion onto the central black hole SGR A* at the scales of the event horizon \citep{ngehtchallenge, Emami2023}. The dynamic time-scales for these observations are very short. Observations of Sgr A* in the sub-mm \citep{Bower2015, Wielgus2022} and near-infrared regime \citep{Gravity2018, Gravity2018b} confirm that Sgr A* is time-varying on timescales as short as 30 minutes. The predicted ISCO period varies between 4 minutes and roughly 30 minutes depending on the spin of the black hole. \citet{Palumbo2019} concluded that a well-sampled baseline coverage on timescales of $\sim30$ minutes is needed to recover the source dynamics.

CLEAN \citep{Hogbom1974} and its many variants \citep{Clark1980, Schwab1984, Wakker1988, Bhatnagar2004, Cornwell2008, Rau2011, Mueller2022b} served the community well for decades, but are recently challenged by forward imaging approaches in the spirit of Regularized Maximum Likelihood (RML) methods \citep{Narayan1986, Wiaux2009, Garsden2015, Ikeda2016, Chael2016, Chael2018, Akiyama2017, Akiyama2017b, eht2019d, Mueller2022} and Bayesian approaches \citep{Arras2019, Arras2020, Broderick2020, Broderick2020b}. Recently we developed new multiresolution tools for performing VLBI imaging \citep{Mueller2022, Mueller2022b}. For these multiscalar approaches we designed special wavelet-based basis functions (difference of Gaussian and difference of spherical Bessel functions) and fitted the basis functions to the uv-coverage. In this way we define smooth basis-functions that are well suited to describe (compress) the recovered image features by encoding information about the uv-coverage itself. Some wavelets are most sensitive to gaps in the uv-coverage while others are most sensitive to covered Fourier coefficients. While the signal from latter ones should be recovered, the signal from former ones are suppressed (effectively avoiding overfitting).

As a byproduct for these multiscalar imaging algorithms, we compute the so called multiresolution support \citep{Mueller2022}: a set of wavelet parameters that are deemed statistically significant to represent the recovered image features. The multiresolution support encodes various information about the recovered image. Firstly, it implements a `support constraint' (where is the emission located in the image?). Secondly, it encodes a `spatial constraint' (which spatial scales are needed to represent the image features at these locations?). Especially the second prior information is determined by the spatial scales that are present in the data, i.e. that are covered by baselines in the observation. We demonstrated in \citet{Mueller2022} that the multiresolution support is a powerful prior information very well suited to refine the imaging procedure. In \citet{Mueller2022} we proposed to add amplitudes and phases to the data terms and remove any regularizer term, but solve the resulting optimization problem by only updating the coefficients in the multi-resolution support. The fit to the observed visibilities improved, but without the addition of spurious artifacts that are typical for overfitting.

Among Stokes I imaging, full polarimetric imaging are of interest for the VLBI community both theoretically \citep{Blandford1977, Hardee2007, Kramer2021} and observationally \citep[among many other e.g.][]{Gomez2011, Hovatta2012, Zamaninsab2014, Gomez2016, Poetzl2021, Ricci2022}, in particular at the event horizon scales \citep{eht2021a, eht2021b}. In polarimetric imaging the recorded data are separated into several polarized subbands and recombined in the four Stokes parameters. Essentially we have four Stokes parameters (I, Q, U, V) and corresponding polarized visibilities. Hence, the problem that we aim to solve for the other three Stokes parameters is the same as for Stokes I: recovering a signal from a sparse measurement of the Fourier coefficients. However, there are some slight differences: while the Stokes I image is necessarily non-negative (and this is used during imaging as a prior), this does not have to be true for Stokes Q, U, and V. Moreover, $I^2 \geq Q^2+U^2+V^2$ applies.

The multiresolution support is a well suited prior to be applied to the polarimetric imaging when the Stokes I image is already done. The `support constraint' of the multiresolution support encodes the information that linear and circular polarized emission theoretically can only appear at locations where total intensity (Stokes I) is bigger than zero. This might not reflect the observation situation in every case: sometimes the Stokes I signature cannot be retrieved with the spatial sensitivity of the interferometer while the more localized (e.g. due to Faraday rotation) polarized structural pattern is visible. However, in most VLBI studies this pathological situation does not appear and `support constraint' is a good approximation. Moreover, the `spatial constraint' adheres the fact that the polarimetric visibilities have the same uv-coverage as total intensity visibilities, i.e. the same spatial scales (the ones covered by uv-coverage) are present in the polarized images.

Another domain of current research is the study of dynamic sources, such as Sgr A*, i.e. the static imaging of a dynamically evolving source as in \citet{eht2022c} and the dynamic movie reconstruction \citep{ngehtchallenge}. In this work we focus on latter problem. Data sets of dynamic sources pose additional challenges. Due to the short variability time scale, the effective uv-coverage in every frame is not sufficient for efficient snapshot imaging. Modern approaches utilize a temporal correlation instead, in a Bayesian framework \citep{Bouman2017, Broderick2022, ngehtchallenge} or as temporal regularizer in the RML framework \citep{Bouman2017, Johnson2017, ehtim, ngehtchallenge}. Moreover, the variability of the source could be misidentified with the calibration of the gains \citep{eht2022c}.

Again the multiresolution support (computed for the time-averaged image) encodes prior information that is very desired for dynamic imaging. The `support constraint' encodes the information that every location of an emission spike appearing during the observation is present also in the mean image. The uv-coverage of the full observation run is the sum of the uv-coverages of the single frames. Hence, the `spatial constraint' also provides some powerful image prior for dynamic imaging: the multiresolution support only allows spatial scales that are present in the mean image (in the full observation run), i.e. the fit in the gaps of the uv-coverage remains under control. On the other hand, the `spatial constraint' allows for the addition of the spatial scales to single frames that might be not represented in the uv-coverage of this single frame, but in earlier or later snapshots. However, we like to mention that there may be a bias towards larger scales since the mean image suppresses small-scale structures present in only part of the individual frames.

Based on the success of the approach presented in \citet{Mueller2022} of only changing the coefficients in the multiresolution support to introduce effective regularization, we propose the same approach for static polarimetric imaging and dynamic imaging. As outlined above, the multiresolution support is well suited to be used as a regularizer in these problems as it exactly encodes the prior information that is needed. As we solve two quite different extensions to the standard VLBI imaging with the same approach, it is natural to use the same approach also for the combined problem: a dynamic, polarimetric reconstruction.

\section{Theory}

\subsection{VLBI}
As described by the van-Cittert-Zernike theorem the visibilities $\mathcal{V}$ are related to the true sky-brightness distribution $I(x,y)$ by a two-dimensional Fourier transform under reasonable assumptions \citep{Thompson2017}:
\begin{align}
    \mathcal{V}_I (u, v) = \int \int e^{-2 \pi i (x u + y v)} I(x, y) dx dy =: \mathcal{F} I (u, v)  \label{eq: vis}.
\end{align}
From a full coverage of the Fourier coefficients (visibilities) the true sky brightness distribution could be computed by an inverse Fourier transform. However, in VLBI the uv-coverage is very sparse with significant gaps. This makes the problem of recovering the image an ill-posed inverse problem. The polarized quantities are measured at every antenna with orthogonal polarimetric filters (linear or circular). The cross-correlation of these signals give rise to the polarimetric Stokes I parameters and their respective polarimetric visibilities:
\begin{align}
    &\mathcal{V}_I =  \mathcal{F}I, \\
    &\mathcal{V}_Q =  \mathcal{F}Q, \\
    &\mathcal{V}_U =  \mathcal{F}U, \\
    &\mathcal{V}_V =  \mathcal{F}V, \\
\end{align}
where $I$ is the total brightness, $Q$ and $U$ the linear polarizations and $V$ the fraction of circular polarization. By construction it is:
\begin{align}
I^2 \geq Q^2+U^2+V^2. \label{eq: side_polarimetry}
\end{align}

\subsection{Imaging}
Imaging with the CLEAN algorithm and its variants \citep{Hogbom1974, Schwab1984, Wakker1988} were the standard in VLBI imaging for the last decades. In CLEAN the imaging problem is equivalently reformulated as a deconvolution problem:
\begin{align}
    I^D = B^D * I,
\end{align}
where $I^D$ is called the dirty map (inverse Fourier transform of all measured, and probably reweighted, Fourier coefficients) and $B^D$ (the dirty map of a synthetic delta source) is called the dirty beam. The astronomer using CLEAN determines some search windows for components, CLEAN looks for the maximum peak in the residual in this window (minor loop) and subtracts the shifted and rescaled dirty beam from the residual (major loop). This procedure is iterated until the residual is noise-like. In this way, CLEAN models the image as a set of delta functions. Finally, these components are restored with a restoring beam (clean beam) that fits the central peak of the dirty beam. CLEAN is an inverse modeling approach to the imaging problem.

Recently forward modeling approaches gained interest in the community in the framework of RML \citep{Chael2018, Akiyama2017b, Mueller2022} and Bayesian methods \citep{Arras2019, Broderick2020, Broderick2020b}. These methods seem to outperform classical CLEAN in terms of speed, spatial resolution, sensitivity and precision, in particular when the uv-coverage is sparse \citep[e.g.][]{eht2019d, Arras2020, Mueller2022, ngehtchallenge}. On the other hand, these forward modeling methods require the fine-tuning of some hyper-parameters and regularization parameters, despite the recent effort to reduce this dependence \citep{Mueller2022}. For the remainder of this manuscript we focus on RML methods and ignore Bayesian approaches for now. 

In RML, a sum of data fidelity terms and penalty terms is minimized:
\begin{align}
    \hat{I} \in argmin_I \sum_i \alpha_i S_i(I) + \sum_j \beta_j R_j(I),
\end{align}
where the data fidelity terms $S_i$ measures the fidelity of the recovered solution $I$ to the observed data (i.e. polarized visibilities) and the penalty terms/regularization terms $R_j$ measure the fidelity of the guess image $I$. The regularization parameters $\alpha_i$ and $\beta_j$ are manually set weights that balance data fidelity and regularization terms. Typical choices for the data terms are chi-squareds to the observed (polarimetric) visibilities, and related calibration independent quantities such as closure phases and closure amplitudes. For the regularization terms a wide range of regularizers has been applied in the past, e.g. sparsity promoting regularization (l1, l2), smoothness constraints (total variation, total squared variation), hard constraints (total flux, non-negativity), entropy maximization (MEM) or multiscale decompositions (hard thresholding on scales). The regularization terms introduce regularization to the ill-posed imaging problem. By balancing the data terms and the regularization terms, we select a possible guess solution that is fitting data (small data terms) and robust against noise and artifacts (small penalty terms). We have demonstrated in previous works \citep{Mueller2022} that a support constraint has the same regularization effect. By constraining the space of free parameters to the multiresolution support we were able to refine the fit to the observed data in later imaging rounds.

\subsection{Wavelets}
The basis behind multiscalar approaches are multiscalar dictionaries. We proposed in \citep{Mueller2022} the use of radial-symmetric difference of Gaussian (DoG) wavelets and extended them to directional dependent basis functions in \citep{Mueller2022b}. Moreover, we introduced in \citep{Mueller2022b} steep, quasi-orthogonal basis functions to study the Fourier domain by difference of Bessel functions (DoB). Both dictionaries (DoG and DoB) are related to each other: the DoG wavelets approximate the central peak of the DoB wavelets, but do not contain the wider sidelobes of latter ones. In what follows we quickly summarize these wavelet dictionaries. For more detailed information we refer to \citep{Mueller2022, Mueller2022b}. 

Wavelets have a wide range of applications in image compression. The most widely used continuous wavelet is the Mexican-hat wavelet which is a rescaled second order derivative of a Gaussian (Lagrangian of Gaussians) \citep{Starck2015}. The difference of Gaussian method offers some viable approximation to Mexican hat wavelets. A DoG-wavelet is described by two width parameters $\sigma_1, \sigma_2$:
\begin{align} \nonumber
    \Phi_\mathrm{DoG}^{\sigma_1, \sigma_2}(x, y) &= \frac{1}{2 \pi \sigma_1^2} \exp \left( \frac{-r(x,y)^2}{2 \sigma_1^2} \right) - \frac{1}{2 \pi \sigma_2^2} \exp \left( \frac{-r(x,y)^2}{2 \sigma_2^2} \right) \\
    &= G_{\sigma_1} - G_{\sigma_2}. \label{eq: dog}
\end{align}
The Fourier transform of these DoG-wavelets define ring-like filters in the Fourier domain:
\begin{align}
    \mathcal{F} \Phi_\mathrm{DoG}^{\sigma_j, \sigma_{j+1}}(u,v) \propto \exp \left( -2 \pi^2 \sigma_j^2 q(u,v)^2  \right) - \exp \left( -2 \pi^2 \sigma_{j+1}^2 q(u,v)^2  \right). \label{eq: fourier_dog}
\end{align}
The extension to DoB-wavelets is natural. We replace the DoG-wavelets, just by spherical Bessel functions:
\begin{align} \nonumber
    &\Phi_\mathrm{DoB}^{\tilde{\sigma}_j, \tilde{\sigma}_{j+1}} (x, y) = \\
    &\frac{1}{\tilde{\sigma}_j r(x, y)} J_1(2 \pi r(x, y) / \tilde{\sigma}_j) - \frac{1}{\tilde{\sigma}_{j+1} r(x, y)} J_1(2 \pi r(x, y) / \tilde{\sigma}_{j+1}).
\end{align}
Moreover, the extension of both wavelets to directional dependent basis functions is straightforward as well. One just has to replace the radial coordinates by elliptical ones. 

The wavelet decomposition is composed out of the wavelet basis functions from a sequence of increasing widths $\sigma_0 \leq \sigma_1 \leq ... \leq \sigma_J$:
\begin{align}
\Psi^\mathrm{DoG}: I \mapsto \mathscr{I} = [\Phi_\mathrm{DoG}^{\sigma_0, \sigma_{1}} * I, \Phi_\mathrm{DoG}^{\sigma_1, \sigma_{2}} * I, ..., G_{\sigma_J} * I], \\
\Psi^\mathrm{DoB}: I \mapsto \mathscr{I} = [\Phi_\mathrm{DoB}^{\sigma_0, \sigma_{1}} * I, \Phi_\mathrm{DoB}^{\sigma_1, \sigma_{2}} * I, ..., J_{\sigma_J} * I].
\end{align}
For direction dependent dictionaries, we use elliptical Gaussians and Bessel functions instead. For more details we refer to our discussion in \citet{Mueller2022b}. The multiscale dictionary is the adjoint of the multiscale decomposition (in what follows called $\Gamma$):
\begin{align}
\Gamma: \mathscr{I} = \{ I_0, I_1, I_2, ..., I_J \} \mapsto \sum_{i=0}^{J-1} \Phi_\mathrm{DoG}^{\sigma_i,\sigma_{i+1}} * I_i + G_{\sigma_J} * I_J,
\end{align}
with an analogous action for DoB-wavelets and multi-directional wavelets. The complete action of the multi-scalar and multi-directional wavelet decomposition is presented in the Appendix.

\subsection{DoG-HiT} \label{sec: doghit}

Our novel algorithm for doing dynamic polarimetric reconstructions is an extension of the DoG-HiT algorithm \citep{Mueller2022b}. We summarize this algorithmic framework in this section. DoG-HiT models the image by a radial symmetric wavelet dictionary $\Psi^\mathrm{DoG}$. The Fourier transform of the basis functions of the dictionary (atoms) are sensitivity filters in the Fourier domain. Hence, by fitting the widths of the Gaussians to the uv-coverage, we define wavelets that are most sensitive to measured Fourier coefficients and wavelets that are most sensitive to gaps in the uv-coverage. The signal of former ones should be kept, while the lack of later atoms causes sidelobes in the image. In this way, the dictionary allows for a better separation between measured features (covered by baselines) and uncovered artifacts. We interpolate the signal in the gaps by the smooth nature of the basis functions, but suppress the signal in the gaps to a level that overfitting is prohibited. All in all, we solve the minimization problem \citep{Mueller2022}:
\begin{align} \nonumber
    \hat{\mathscr{I}} \in \mathrm{argmin}_\mathscr{I} & \left[  S_\mathrm{cph}(F \Gamma \mathscr{I}, \mathcal{V}) + S_\mathrm{cla}(F \Gamma \mathscr{I}, \mathcal{V}) \right. \\
    &\left. + \beta \cdot \norm{\mathscr{I}}_{l^0} + R_\mathrm{flux} (\mathscr{I}, f) \right], \label{eq: second_round_problem}
\end{align}
where $S_\mathrm{cph}$ and $S_\mathrm{lca}$ denote the $\chi^2$-fit to the closure phases and closure amplitudes respectively. $R_\mathrm{flux}$ denotes a characteristic function on the total flux of the guess solution. We use the pseudo-norm $\norm{\cdot}_{l^0}$(i.e. the number of non-zero coefficients) as a sparsity promoting regularization term weighted with a regularization parameter $\beta$. Eq. \eqref{eq: second_round_problem} is solved by a forward-backward splitting algorithm alternated with rescaling the emission to a predefined total flux \citep{Mueller2022}. The final recovered solution is:
\begin{align}
\hat{I} = \Gamma \mathscr{I}.
\end{align}
The regularization parameter $\beta$ is the only free parameter that needs to be chosen manually by the user. The number of free parameters is therefore much smaller than the number of free parameters for RML methods such as ehtim \citep{Chael2016, Chael2018} or SMILI \citep{Akiyama2017, Akiyama2017b} since the penalty term is chosen data-driven. We demonstrated in \citet{Mueller2022} that although the optimization landscape is much simpler, the reconstructions obtained by DoG-HiT are competitive to RML reconstructions.
Moreover, we only fit closure phases and closure amplitudes for DoG-HiT in Eq. \eqref{eq: second_round_problem}, i.e. the reconstruction is robust against instrumental gain corruptions. Consecutively we use the model computed by DoG-HiT for self-calibration, i.e. we determine the gains.

\subsection{Multiresolution support}
\begin{figure*}
    \hspace{-3cm}
     \begin{subfigure}[b]{0.45\textwidth}
         \centering
         \includegraphics[width=\textwidth]{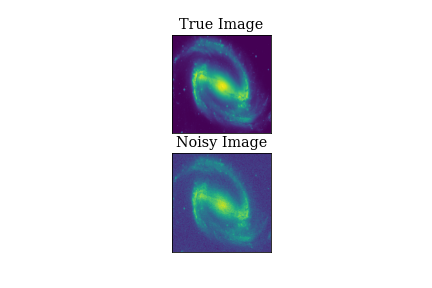}
         \label{fig:y equals x}
     \end{subfigure}
     \hspace{-2.8cm}
    \begin{subfigure}[b]{0.45\textwidth}
         \centering
         \includegraphics[width=\textwidth]{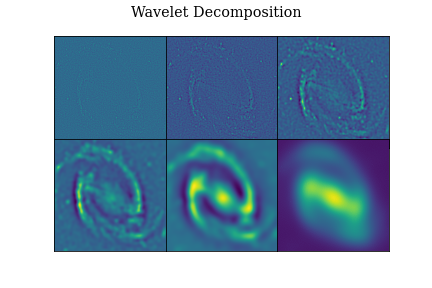}
         \label{fig:y equals x}
     \end{subfigure}
     \hspace{-1cm}
    \begin{subfigure}[b]{0.45\textwidth}
         \centering
         \includegraphics[width=\textwidth]{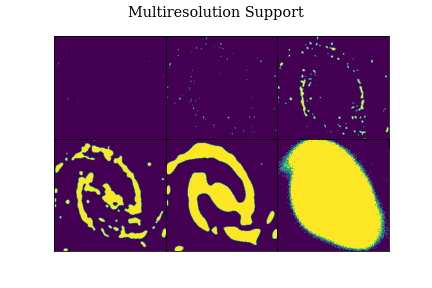}
         \label{fig:y equals x}
     \end{subfigure}     
    \caption{Left panels: true image and true image with additional Gaussian noise, middle panels: wavelet decomposition of the noised image with the DoG-wavelet dictionary computed with filter sizes $\sigma_0=1, \sigma_1=2, \sigma_2=4, ..., \sigma_5 = 32$ pixels, right panels: multiresolution support computed by thresholding the wavelet scales to the scale-dependent noise plotted as a mask with either value 1 (coefficient in the support) or 0 (coefficient not in the multiresolution support)}.
    \label{fig: wavelet_decomposition}
\end{figure*}

A specific property of the multiscalar decompositions is the multiresolution support. \citep{Mertens2015} paved the way for the application of the multiresolution support in the analysis of AGN jets. The multiresolution support is a set of wavelet components that are statistically significant \citep{Starck2015}. We decompose a noisy image by a wavelet dictionary: $[I_0, I_1, I_2, ..., I_J] = \Psi I$. Moreover, we compute the scale-dependent noise-level $s_j$ by decomposing a Gaussian white noise field with the same wavelet dictionary. Given some threshold $k_s$, we can define a set of statistically significant wavelet coefficients with the criterion that $\norm{I_j(x,y)} \geq k_s s_j$ where the noise-level is approximated by the variance from an emission-free region of the image scale $I_j$ (i.e. far away from the center). The multiresolution support for a celestial ground truth image from the EHT imaging challenges \footnote{http://vlbiimaging.csail.mit.edu/} is illustrated in Fig. \ref{fig: wavelet_decomposition}. 

The multiresolution support encodes two different types of prior information about the model. Firstly, it encodes a `support constraint', i.e. it defines the position of significant emission spikes in the field of view. 

Secondly, the multiresolution support contains information about the spatial scales that are present in the observation. In sparse VLBI arrays, this is dominated by the uv-coverage, i.e. by which spatial scales are covered by observed baselines in the Fourier domain. As various wavelet basis functions are most sensitive to various baselines or gaps in the uv-coverage, the information about which spatial scales are covered by observations is directly imprinted in the multiresolution support. This is especially true for the direction dependent DoG- and DoB-wavelets used for DoG-HiT that were fitted to the uv-coverage, i.e. that were developed to allow an optimal separation between covered features and gaps in the uv-coverage.

DoG-HiT solves the minimization of Eq. \eqref{eq: second_round_problem} with a forward-backward splitting algorithm. The backward projection step is the application of the proximal-point operator of the $l^0$ penalization function, which is a hard thresholding \citep{Mueller2022}. Hence, all insignificant wavelet coefficients are set to zero. DoG-HiT therefore computes an approximation of the multiresolution support as a byproduct. This support was used for further refining rounds in the imaging \citep{Mueller2022}.

The computation of the multiresolution support as a byproduct of DoG-HiT highlights an essential improvement of DoG-HiT compared to CLEAN regarding supervision. The support of significant emission is found by DoG-HiT automatically, while it has to be selected in CLEAN by the user-defined CLEAN windows. DoG-HiT is therefore is less user-biased and provides (compared to standard RML frameworks and CLEAN) an essential step towards unsupervised VLBI imaging.

\section{Algorithms}
We outline in this section the algorithms used for static polarimetry, dynamic Stokes I imaging and dynamic polarimetry. In what follows, we will call these algorithms `mr-support imaging'.

\subsection{Stokes I} \label{sec: static_I}
Static Stokes I images are constructed with DoG-HiT with the five round pipeline presented in \citep{Mueller2022}. However, in \citep{Mueller2022} we used only radially symmetric wavelets. As an extension, we use the multi-directional dictionaries developed in \citep{Mueller2022b} in this work, i.e. we replace the circular symmetric Gaussians by elliptical Gaussians. Moreover, we used a grid search in \citep{Mueller2022} to find a proper starting point for the forward-backward splitting minimization iterations of DoG-HiT. Since the backward step in the minimization is essentially a hard thresholding, we tried different scale-dependent thresholds in an equidistant grid to minimize Eq. \eqref{eq: second_round_problem} and used the setting of the minimum as the starting point for the forward-backward iterations. For this manuscript, we use the same grid search, but apply the orthogonal DoB-wavelets in the grid search, while still using the DoG wavelets in the imaging rounds of the pipeline. We will not focus on the Stokes I reconstruction in this work as these extensions are rather straightforward and minor, and the focus of the manuscript is on an extension of DoG-HiT to polarimetry. We recall one of the main advantages of DoG-HiT: the algorithm works mainly unsupervised with a minimal set of free parameters, hence adding a minimal human bias in the imaging procedure.

\subsection{Polarimetry} \label{sec: static_pol}
For polarimetric reconstructions we first reconstruct a Stokes I image with DoG-HiT and solve for the gains by self-calibrating to the final output (note that DoG-HiT relies on calibration independent closure quantities). As a second step, we solve for the polarimetric Stokes parameters $Q,U$ and $V$. We take the multiresolution support computed by DoG-HiT for the Stokes I imaging and constrain the space of free parameters to all wavelet coefficients in the multiresolution support. We then solve for $Q,U,V$ by minimizing the fit to $\mathcal{V}_Q, \mathcal{V}_U, \mathcal{V}_V$ with a gradient descent algorithm, but only allow coefficients in the multiresolution support to vary. In summary we solve the following problems:
\begin{align} \nonumber
& \hat{\mathcal{Q}} \in \mathrm{argmin}_{\mathcal{Q} = \{Q_0, ..., Q_n \}, Q_j(x,y)=0\,whenever\,\hat{I}_j(x,y)=0} \left[ S_Q(F \Gamma \mathcal{Q}, \mathcal{V}_Q) \right] \\
& \hat{\mathcal{U}} \in \mathrm{argmin}_{\mathcal{U} = \{U_0, ..., U_n \}, U_j(x,y)=0\,whenever\,\hat{I}_j(x,y)=0} \left[ S_U(F \Gamma \mathcal{U}, \mathcal{V}_U) \right], \label{eq: constrainedmin}
\end{align}
where $\{\hat{I}_0, ..., \hat{I}_n\} =: \hat{\mathscr{I}}$ are the recovered wavelet coefficients for the Stokes I image as in Sec. \ref{sec: doghit}. $S_U$ and $S_Q$ are the $chi^2$-fit qualities to the Stokes Q and U visibilities. The side condition $Q_j(x,y)=0$ whenever $\hat{I}_j(x,y)=0$ denotes the constraint that we only vary coefficients in the multiresolution support.

The multiresolution support is a well suited regularizer here: the support constraint encodes the side-condition Eq. \eqref{eq: side_polarimetry} effectively, i.e. polarized emission is only allowed to appear at locations in the images in which we found relevant emission in total intensity. While this inequality \eqref{eq: side_polarimetry} holds true theoretically in any case, in practice the pathological situation could occur that due to the instrumental effect a non-detection of Stokes I does not rule out polarimetric structures. With this caveat in mind, we assume for the rest of the manuscript that inequality \eqref{eq: side_polarimetry} holds true in observations as well. Moreover, the polarimetric visibilities have the same uv-coverage as the Stokes I visibility. The `spatial constraint' of the multiresolution support describes which spatial scales are statistically significant to describe the emission in the image, which in case of sparse VLBI arrays is dominated by the uv-coverage (i.e. which spatial scales are compressed by which baselines and whether these baselines are measured). Hence, we already computed the multiresolution support as a byproduct in DoG-HiT to study the uv-coverage of the observation and get control over overfitting in the gaps of the uv-coverage by suppressing the respective atoms of the dictionary. This effective regularization can be copied over to the polarized visibilities as the uv-coverage is the same.

Moreover, we like to stress out once again that the multiresolution support is a completely data driven property computed as a sideproduct by DoG-HiT. Hence, the reconstruction of polarimetric properties still relies on a minimal set of hyper-parameters and remains largely unsupervised.

We fit complex polarimetric visiblities directly here. That requires that a good polarization calibration is available already. The method is however easy to adapt to more realistic situations since it is (opposed to CLEAN) a forward-modeling technique. Firstly, instead of a constrained $\chi^2$-minimization to the complex visibilities, one could just optimize the fit to the visibility-domain polarization fraction as in \citep{Johnson2015}. Secondly, the minimization in Eq. \eqref{eq: constrainedmin} is done iteratively, where the most important features are recovered first and gradually more detailed features will be recovered at later iterations. Hence, with a similar philosophy to how self-calibration interacts with CLEAN, we could run the minimization for some iterations and do the calibration on the current model, then continue the minimization and calibration in an alternating manner.

\subsection{Dynamic Stokes I} \label{sec: dynamic_I}
For dynamic Stokes I imaging, we first reconstruct a static image with DoG-HiT. For this work we assume that the static image of a dynamically evolving source might be a good approximation to the mean image during the time of observation. This might be in particular true if the source structure contains some persistent structure during the complete observing run, as could be expected for Sgr A* in EHT observations with a persistent shadow in rotating hotspot models \citep{Tiede2020b}. However, based on the dynamics of the target, it may be difficult to recover a decent fit to the data with a static image. In this work we applied a procedure inspired by the strategy in \citet{eht2022c}, i.e. we added a systematic noise-floor on every baseline to account for variability. However, we did not repeat the sophisticated noise modeling applied in \citet{eht2022c}.

We compute the multiresolution support by the static mean image. Then, we cut the observation in single frames and reconstruct images at every frame independently. All frames together make up the dynamic movie reconstruction. However, due to the shortness of single frames, snapshot imaging is not possible due to the sparsity of the uv-coverage. Again we propose to use the multiresolution support instead. We minimize the $\chi^2$ for every single frame observation independently for every frame in a gradient descent algorithm (using the mean image as an initial guess), but only allow coefficients in the multiresolution support to vary. 

The multiresolution support is a well suited regularizer here as well: if the static image is a good approximation to the mean image, the static image contains all the locations of emission in the field of view. If at some time an emission spike occurs at a specific location, this emission spike should be visible in the mean as well. Hence, the `support constraint' encodes information about the location of emission at single frames. This assumption comes with the caveat that short-living, small-scale features may be not strong enough in the mean image and excluded later from the dynamic reconstructions due to the multiresolution support. However, we also doubt that such a feature would be visible with the much sparser uv-coverage of single scans, and therefore would not be recovered anyways. Moreover, the uv-coverage of the complete observation is the sum of the observations of the single frames. In single frame observations there are three different categories of Fourier coefficients/baselines: the ones measured by observations in this single frame (very sparse), the ones that are not measured during the time of the single frame, but will be measured at later (earlier) times in the observation, and the baselines that are not measured at all due to the sparsity of the array. By doing constrained optimization (constrained by the multiresolution support) to the single frame observation we fit the first class of baselines, copy the solution over from the initial guess (mean image) for the second class of baselines, and suppress the last class of baselines by the multiresolution support. Hence, the `spatial constraint' implemented by the multiresolution support is a well suited prior to do dynamic imaging.

The reasonable assumption of temporal correlation between scans, e.g. by a regularizer term favoring temporal smoothness, is not used explicitly for mr-support imaging. However, such assumptions could be included in the dynamic reconstruction straight-forwardly: instead of fitting the visibilities with a constrained minimization approach, we minimize the sum of a quality metric for the fit to the visibilities and a temporal regularization term, but only vary coefficients in the multiresolution support. However, for this work we restrict ourselves to reconstructions without penalization on the temporal evolution such that now new regularization parameters are introduced and the reconstruction remains automatic and completely data-driven. Moreover, due to this fact all scans can be computed in parallel allowing for fast computations.

\subsection{Dynamic polarimetry} \label{sec: dynamic_pol}
We propose the same procedure for polarized imaging and dynamic Stokes I imaging: fitting the respective visibilities with a gradient descent approach while only varying coefficients in the multiresolution support computed by DoG-HiT. It is therefore natural to utilize this approach for dynamic polarimetry as well. In fact, we propose the following strategy. First reconstruct a static Stokes I image by DoG-HiT and compute the multiresolution support. Then cut the observation in single frames and solve for dynamics and polarimetry together by fitting to $\mathcal{V}_I, \mathcal{V}_Q, \mathcal{V}_U, \mathcal{V}_V$ together in single frames independently, but only vary coefficients in the multiresolution support.

\section{Synthetic data tests} \label{sec: tests}

\subsection{Synthetic observations} \label{sec: observation}
We tested the capabilities for mr-support imaging for polarimetric image reconstructions. We test three different source models (static polarized Sgr A* model, a slowly rotating crescent and a rapidly rotating crescent) with two different arrays (EHT and a possible ngEHT configuration). A thorough comparison of existing imaging approaches for dynamic polarimetry is in preparation and will be deferred to a consecutive work. For more details we also refer to the ngEHT analysis challenges \citet{ngehtchallenge}, and in particular the upcoming third challenge \footnote{https://challenge.ngeht.org/challenge3/} in which we compete with mr-support imaging. We review our submission to the third challenge in Sec. \ref{sec: challenge}.

We observe the synthetic ground truth images and movies with the array of the EHT 2022 observations and added thermal noise according to the measured SEFDs of the 2017 observation campaign \citep{eht2019d}. We used ten minute cycles of five minutes of continued observation with an integration time of ten seconds and a five minute off-source gap (mimicking calibration, pointing scans). This cycle time is of special interest when discussing dynamic reconstructions as the five-minute gaps essentially limit the temporal resolution. The data sets were scan-averaged prior to the imaging procedure.

As ngEHT configuration we took the EHT 2022 array configuration (i.e. ALMA, APEX, GLT, IRAM-30 m, JCMT, KP, LMT, NOEMA, SMA, SMT, SPT) and added ten additional antennas from the list of \citep{Raymond2021} as was done for the ngEHT Analysis challenges \citep{ngehtchallenge}: HAY (34\,m), OVRO (10.4\,m), GAM (15\,m), BAR, BAJA, NZ, SGO, CAT, GARS, CNI (all 6\,m). We added instrumental noise according to the size of the telescopes, but did not add further calibration errors. As a ground truth we took the slowly rotating crescent model with a rotation period of one hour. As for the EHT 2022 coverage, the ground truth movie is observed with a cycle of five minutes on source and a five minutes gap and an integration time of ten seconds (10 minutes on source and 2 minutes gaps in the fastly rotating crescent example). 

\begin{figure*}
    \centering
    \includegraphics[width=\textwidth]{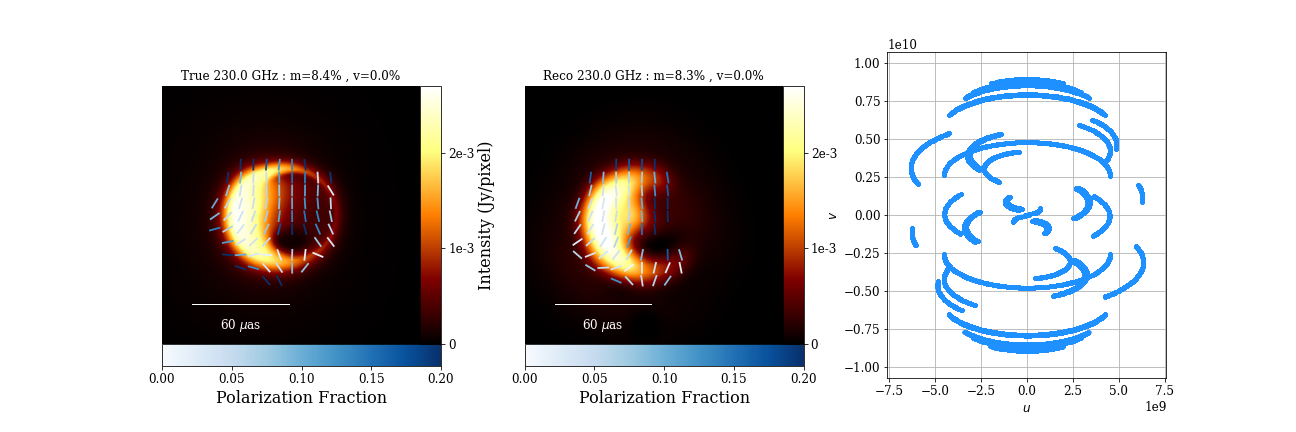}
    \caption{Left panel: static polarization ground truth, middle panel: static reconstruction with mr-support imaging, right panel: uv-coverage of synthetic observation (EHT 2022 array).}
    \label{fig: static_pol}
\end{figure*}

As a static synthetic test image we took a synthetic Sgr A* image out of the \textit{ehtim} software package \citep{Chael2018}. The true image model is presented in Fig. \ref{fig: static_pol}.

\begin{figure*}
    \centering
    \includegraphics[width=\textwidth]{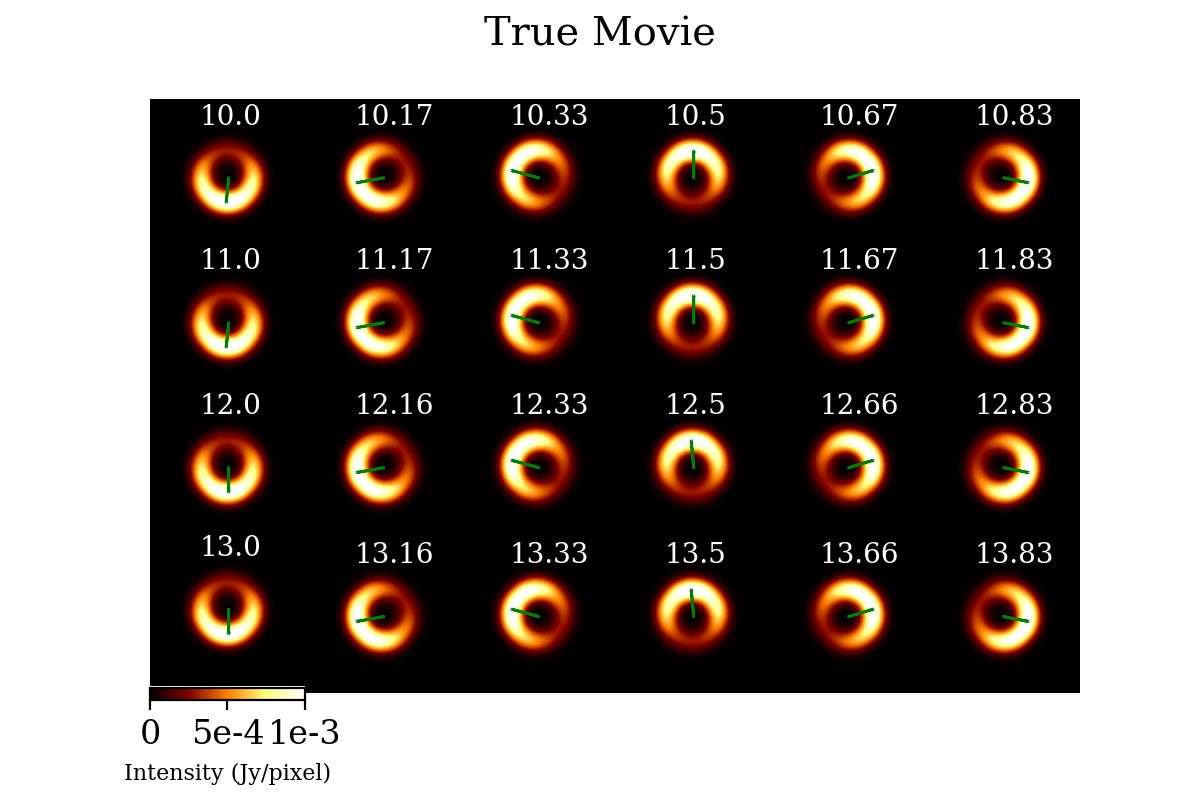}
    \caption{Synthetic ground truth dynamic movie (slowly rotating crescent) in the time interval between 10\,UT and 14\,UT. The green arrow ranges from the image center to the position of the brightest pixel in the frame, hence illustrating the orientation of the crescent.}
    \label{fig: dynamic_true}
\end{figure*}

For the dynamic Stokes I imaging we used a crescent model \citep{Tiede2020}:
\begin{align}
    I(r, \theta) = I_0 (1-s \cos( \theta - \xi)) \frac{\delta(r-r_0)}{2 \pi r_0}.
\end{align}
We use the parameters: $I_0 = 0.6\,\mathrm{Jy}$, $s=0.46$, and $r_0 = 22\,\mu\mathrm{as}$. To account for dynamics roughly similar to rotating hotspot models \citep{Tiede2020b} we let the crescent rotate clockwise. One rotation period takes 1 hour which is roughly comparable to the flux variability time-scale of the SGR A* lightcurve \citep{Wielgus2022}. The synthetic ground truth image is presented in Fig. \ref{fig: dynamic_true}. To illustrate the orientation of the crescent, we also show a green arrow from the image center to the location of the brightest pixel in the image in Fig. \ref{fig: dynamic_true}. For polarized movies we have to add polarization. For the sake of simplicity we used a simpler model for testing the capabilities of dynamic polarimetry here: we added a constant linear polarized structure at $10\%$ (no circular polarization) with a rotating EVPA. To separate the dynamic polarimetric reconstruction from effects of the Stokes I imaging, the rotation of the EVPAs is counter-clockwise (rotation of Stokes I was clock-wise) and has a different rotation period of two hours instead of one hour as for the Stokes I images. 

\begin{figure*}
    \centering
    \includegraphics[width=\textwidth]{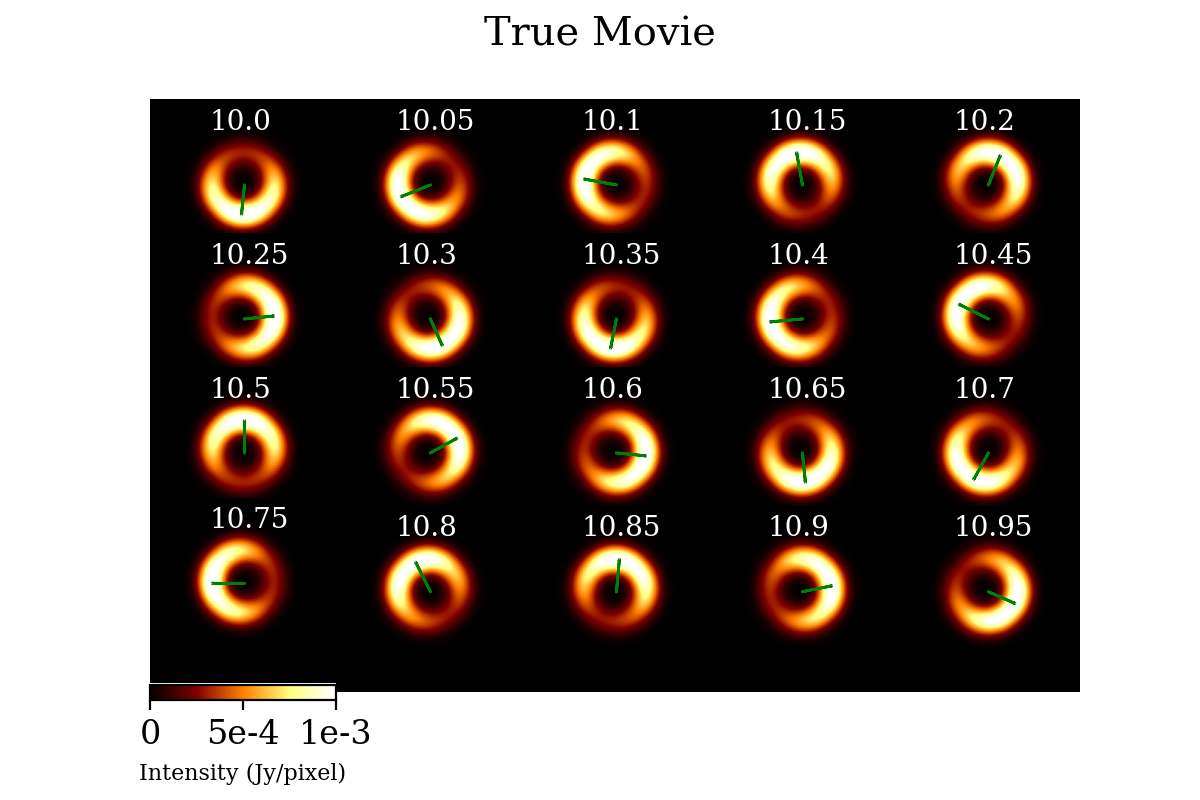}
    \caption{True movie for fast rotating crescent.}
    \label{fig: dynamic_true_ngeht_10_2}
\end{figure*}

As an additional model we also test a rapidly rotating crescent model with an orbital period time of twenty minutes. We show the ground truth movie in Fig. \ref{fig: dynamic_true_ngeht_10_2}. The constant EVPA pattern rotates counter-clockwise in one hour. The advance time between scans that is used for pointing and calibration limits the temporal resolution. For an array as sensitive such as the ngEHT a smaller gap time might be possible. We therefore synthetically observed the rapidly rotating movie with a cycle of ten minutes of scientific observation (ten seconds integration time) and two minutes gaps.

\subsection{Static polarization with EHT coverage} \label{sec: static_I_test}
\begin{figure*}
    \centering
    \includegraphics[width=\textwidth]{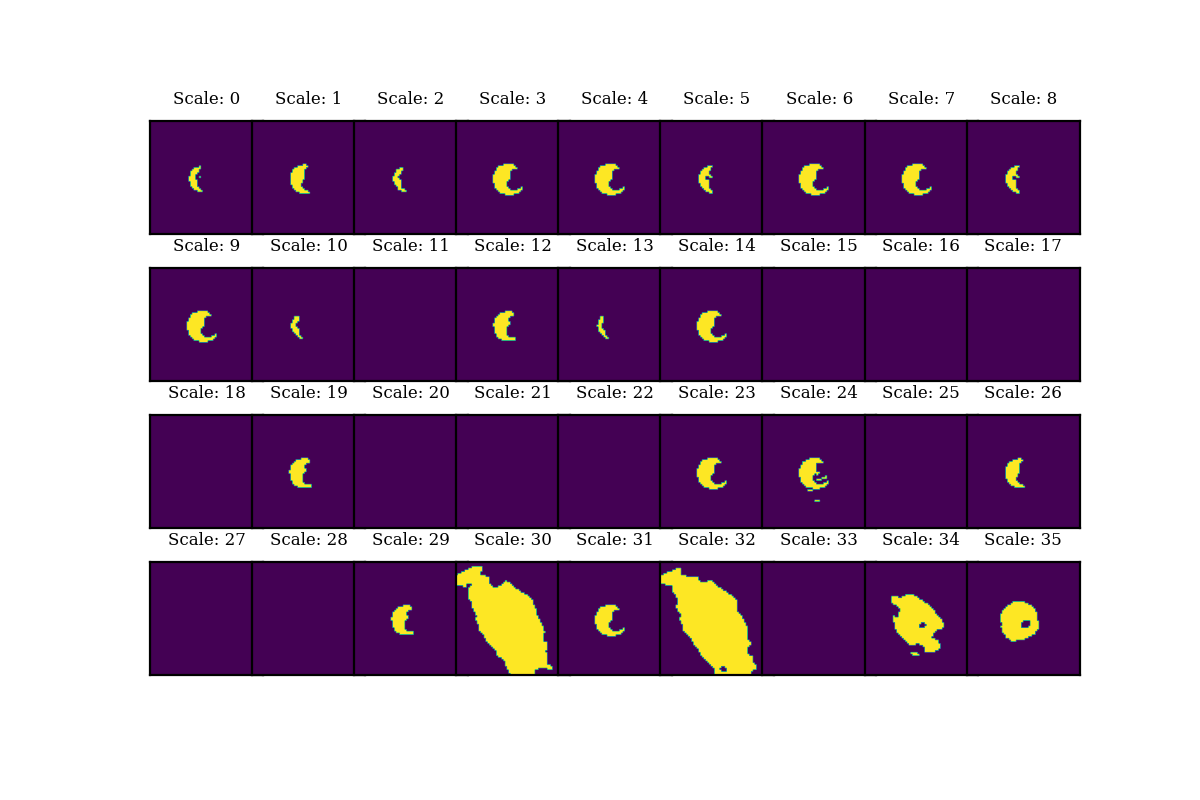}
    \caption{Multiresolution support for the reconstruction of the static polarization example with EHT coverage.}
    \label{fig: support}
\end{figure*}

We fitted the scales to the uv-coverage first with the procedure outlined in \citet{Mueller2022} and \citet{Mueller2022b}: we searched for jumps in the sorted distribution of uv-distances that exceed some threshold and we selected the radial scales accordingly. We defined nine radial scales and used four different angles, resulting in 36 scales to represent the uv-coverage. The Stokes I image was recovered with DoG-HiT \citep{Mueller2022} using the multi-directional dictionaries introduced in \citep{Mueller2022b} as described in Sec. \ref{sec: static_I}. As presented in Sec. \ref{sec: static_pol} we then computed the multiresolution support. The multiresolution support is presented in Fig. \ref{fig: support}. Some scales that are most sensitive to gaps in the uv-coverage are suppressed completely, while other scales encode various parts of the emission structure, i.e. the ring like emission (scale 34 and scale 35), the extended emission structure (scale 30 and 32), the fine crescent structure (among others scale 4, 7, 9, 14 and 24), or the bright spot to the left of the crescent (e.g. scale 0, 2 and 10). The minimization to the polarized visibilities was done with the limited-memory Broyden–Fletcher–Goldfarb–Shanno (BFGS) algorithm \citep{BFGS}, as implemented in {\tt Scipy} \citep{Jones2001}. To assert global convergence, we blurred the Stokes Q and U image of the reconstruction with the nominal resolution and redo the minimization with a gradient descent procedure. 

We show the final reconstruction result in Fig. \ref{fig: static_pol}. The reconstruction of the Stokes I image is relatively successful. The crescent-like shadow image is overall well recovered. However, there are some finer structures that are not recovered by DoG-HiT: the closing of the ring by a thin line towards the right and the fainter structure inside the ring. The linear polarized emission is overall very well recovered. The total fraction of linear polarized light and the overall direction of the electromagnetic vector position angles (EVPA) in North-South direction are well recovered. The synthetic ground truth image contains some more complex, local structures, e.g. a rotation of the EVPA in the bottom left of the image towards an east-west direction. This shift is partly visible in the recovered image as well, although the amount of rotation is smaller.

All in all, this example demonstrates that even for a very challenging and sparse array such as the EHT 2022 array the polarimetric reconstruction with support imaging is quite successful in both the overall structure, but also in the reconstruction of more localized polarimetric structures with a size of $\approx 5\,\mu\text{as}$. Thus, similar to the DoG-HiT reconstruction for the Stokes I image, mr-support polarimetry seems to offer mild super-resolution. Interestingly, super-resolution and a good fit to the polarized visibilities is offered without introducing artifacts in the image. This demonstrates the power of the regularization approach. 

\subsection{Dynamic Stokes I} \label{sec: dynamic_I_test}

The synthetic slowly rotating crescent movie was observed as described in Sec. \ref{sec: observation} with a ten-minute cycle with EHT coverage. According to this temporal resolution, we cutted the observation into frames with a length of ten minutes for the dynamic reconstruction. The reconstruction was then done with the mr-support approach in the best time window $t \in [10\,\mathrm{UT}, 14\,\mathrm{UT}]$ \citep{Farah2022} as outlined in Sec. \ref{sec: dynamic_I}: as a first step we fitted a symmetric ring model to the data, created a mean image with DoG-HiT with the fitted ring model as an initial guess, then we solved sequentially for every frame by mr-support imaging with the support calculated from the mean. As an initial guess for the single frame imaging with mr-support imaging we used the reconstruction of the respectively preceding frame (or the mean in case of the first frame).  

\begin{figure*}
    \centering
    \includegraphics[width=\textwidth]{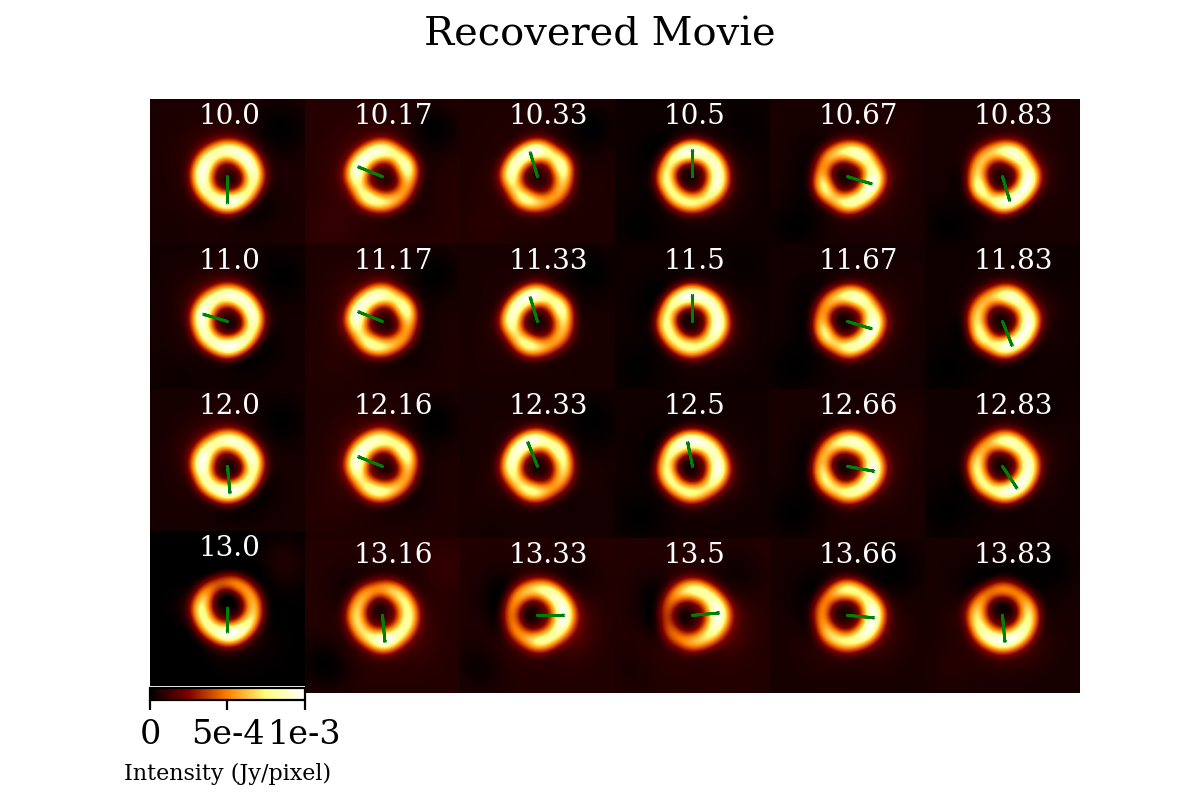}
    \caption{The recovered solution (recovered with mr-support imaging) for slowly rotating crescent observed with the EHT.}
    \label{fig: dynamic_reco}
\end{figure*}

We present the reconstruction results in Fig. \ref{fig: dynamic_reco}. The single frames all show a circular structure with a radius of $\approx 22\,\mu\mathrm{as}$. Moreover, nearly all frames have an asymmetry of a crescent. However, the crescent asymmetry is less prominent than in the true image. As for the true dynamic movie, we illustrate the orientation of the crescent by an arrow from the center to the brightest pixel in the reconstruction. Following the orientations of the recovered crescents in Fig. \ref{fig: dynamic_reco} a clear rotation with an orbital period of one hour is visible. The orientation of the recovered crescents match in most frames with the synthetic ground truth except for some notable exceptions at 11\,UT (no asymmetry recovered at all), and 13.16\,UT-13.5\,UT (wrong orientations). In particular the latter one could be a consequence of taking the reconstruction at the preceding frame as an initial guess for the next frame. The false-recovery at 13.16\,UT hence also affects all following frames. 

\begin{figure}
	\centering
	\includegraphics[width=0.5\textwidth]{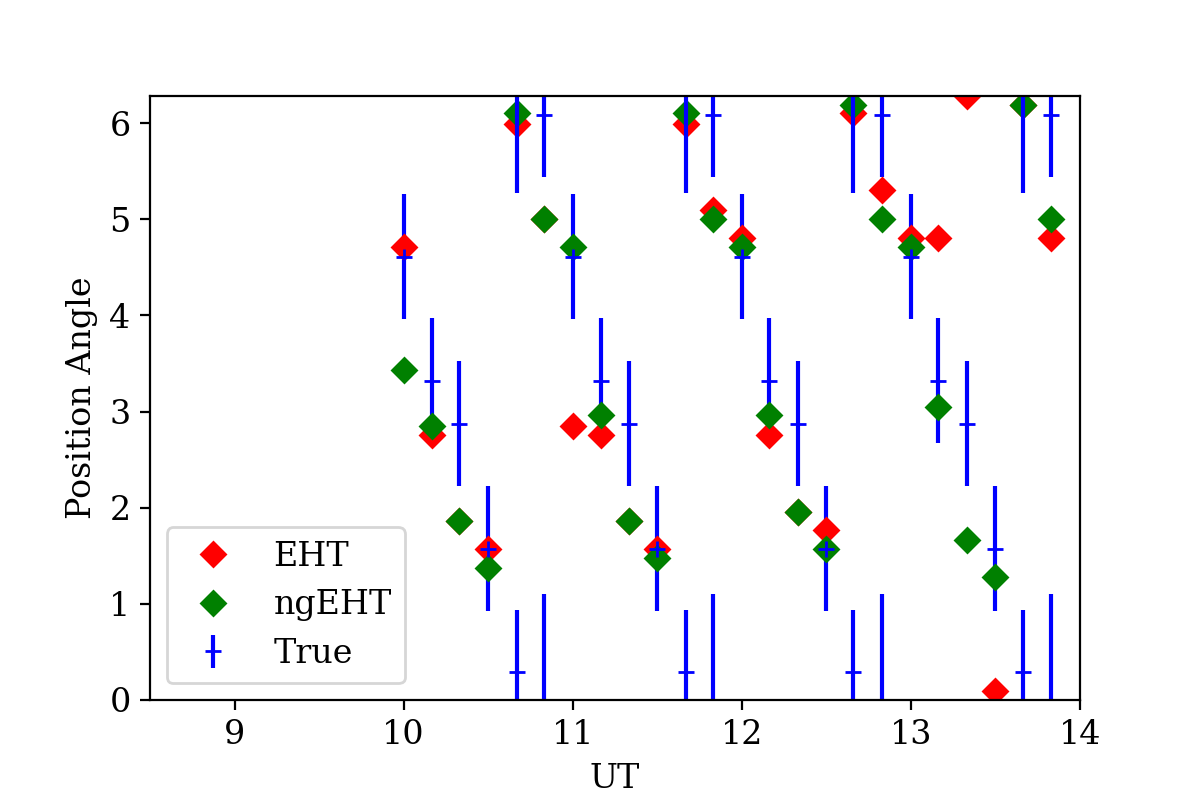}
	\caption{True position angle (blue) and the recovered position angle recovered with mr-support imaging for an EHT configuration (red) and ngEHT configuration (green) for the slowly rotating crescent model. The errorbars reflect the change in position angle in the true source model within a 10 minutes scan (cycle length of synthetic observation).}
	\label{fig: position_angles}
\end{figure}

\begin{figure*}
    \centering
    \includegraphics[width=\textwidth]{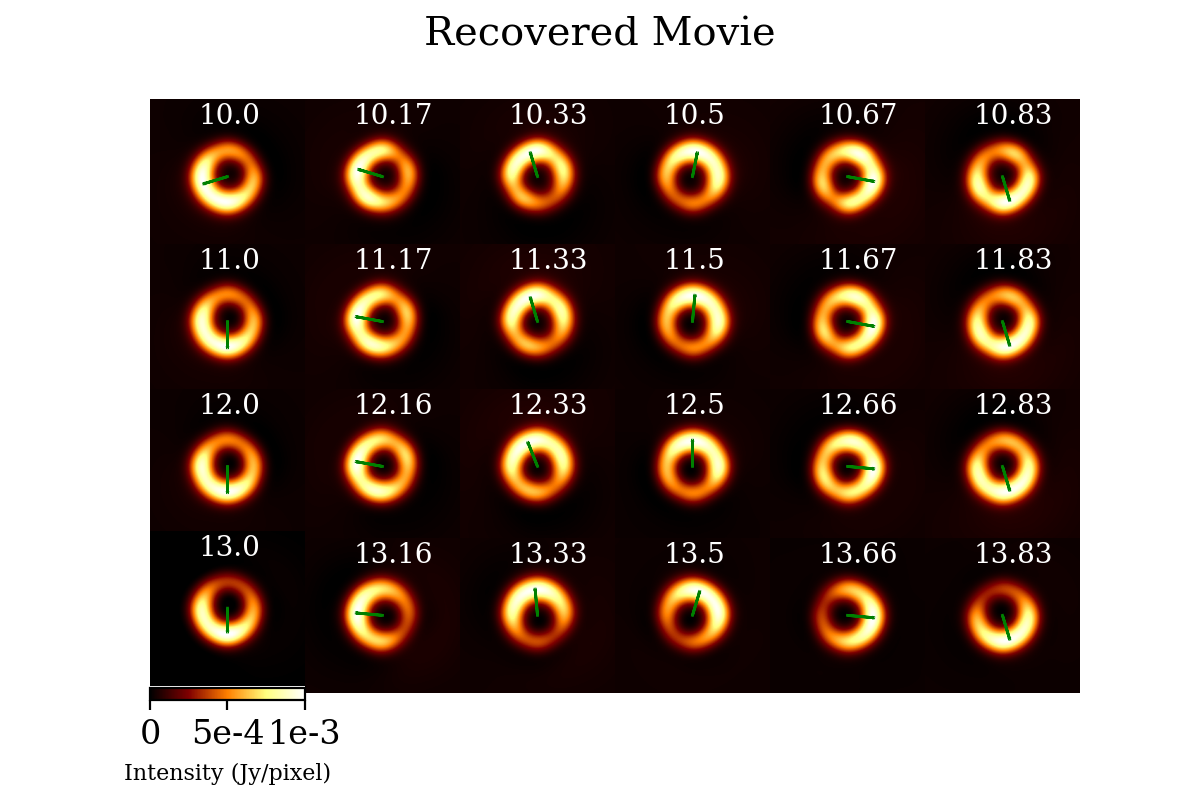}
    \caption{Same as Fig. \ref{fig: dynamic_reco} but with ngEHT coverage: slowly rotating crescent observed with the ngEHT.}
    \label{fig: dynamic_reco_ngeht}
\end{figure*}

We present in Fig. \ref{fig: dynamic_reco_ngeht} the  reconstruction result for a slowly rotating crescent with ngEHT coverage. The reconstruction of the crescent is excellent at every frame with high contrast images. The single-frame images do not show additional image artifacts. Although the additional ngEHT antennas have rather large thermal noise-levels, the much improved density of the array effectively stabilizes against thermal noise. Strikingly the orientation of the crescents matches the ground truth very well. We present in Fig. \ref{fig: position_angles} a comparison between the true position angles and the recovered ones with an error by the temporal smearing due to the scan-length.

\begin{figure*}
    \centering
    \includegraphics[width=\textwidth]{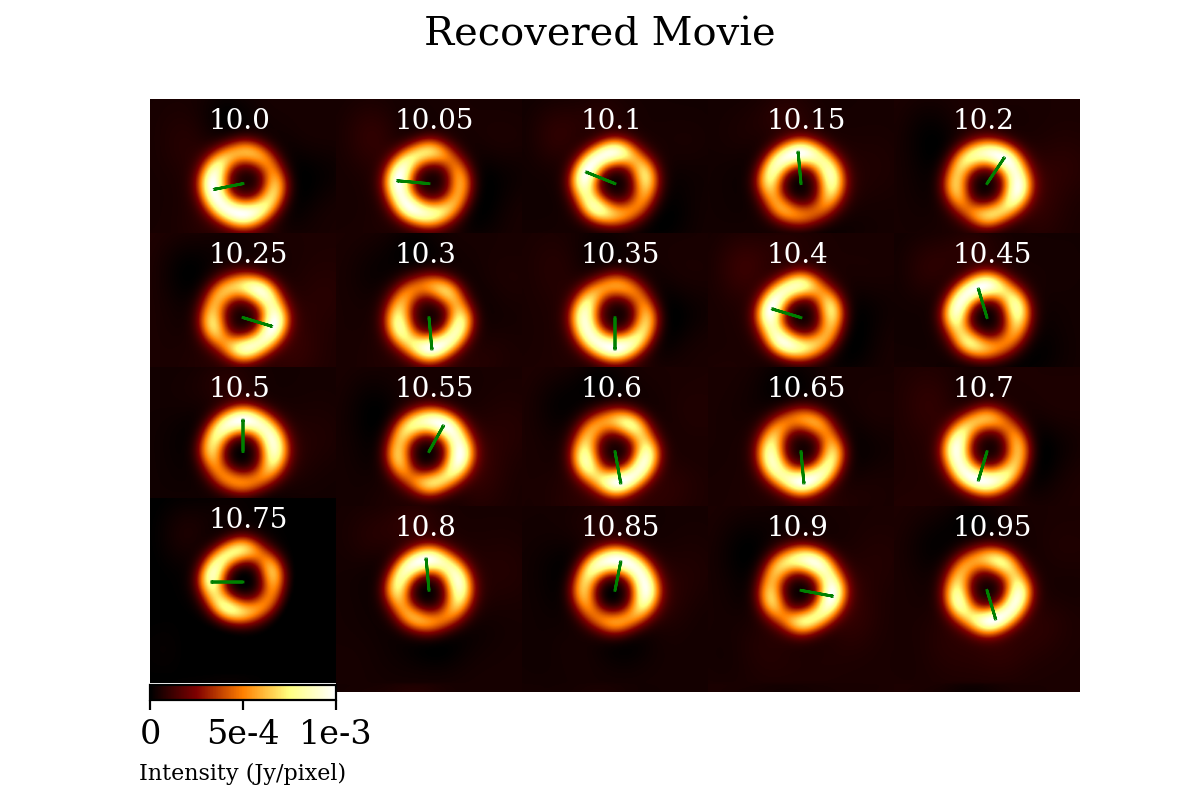}
    \caption{Reconstruction of fast rotating crescent with ngEHT coverage: rapidly rotating crescent observed with the ngEHT.}
    \label{fig: dynamic_reco_ngeht_10_2}
\end{figure*}

The ngEHT array is much denser than the EHT configuration of 2022. This enhances the possible temporal resolutions. We therefore also studied the possibility to observe faster rotating structures at the event horizon with the fast rotating crescent model. The dynamic reconstruction was done in this case in frames of three minutes in length. The faster orbital period and the shorter frame length complicate the reconstruction procedure: there are fewer observation points per single frame which raises the problem of sparsity. Moreover, due to the shorter dynamical timescale and the smaller number of observing points per single frame, the scan-averaged visibility points worsen the signal to noise ratio by a factor of $\sqrt{3}$ compared to the slower rotating crescent. The reconstruction results for dynamic Stokes I imaging with mr-support imaging are shown in Fig. \ref{fig: dynamic_reco_ngeht_10_2}. The crescent is observed at every frame. Additionally the overall orientation matches quite well. However, the quality of the reconstruction decreases compared to the slowly rotating crescent, as can be expected: the asymmetry of the crescents is less clear and the orientation is slightly off by roughly fifteen degrees in some frames. 

All in all, we observe that with mr-support imaging we recover the correct image structure, including overall shadow feature, crescent asymmetry, and orientation, for most frames in the observation very well. Again we like to mention that these particular successful reconstructions do not suffer from introducing image artifacts despite the sparsity of the uv-coverage, especially in single frame observations. This, once again, demonstrates the regularizing property of the mr-support approach.

\subsection{Dynamic polarimetry}
As outlined in Sec. \ref{sec: dynamic_pol} we did the dynamic reconstruction of the Stokes I channel first. Hence, we copied over the reconstructions from Sec. \ref{sec: dynamic_I_test}. We then added polarization frame by frame by mr-support imaging. Similar to our procedure presented in Sec. \ref{sec: static_I_test} we first minimized the data terms (fit to polarized visibilities) with a BFGS minimization procedure, blurred the reconstructed polarized images with the nominal resolution, and minimized the fit with a gradient descent procedure starting from the blurred image as an initial guess. 

\begin{figure*}
    \centering
    \includegraphics[width=\textwidth]{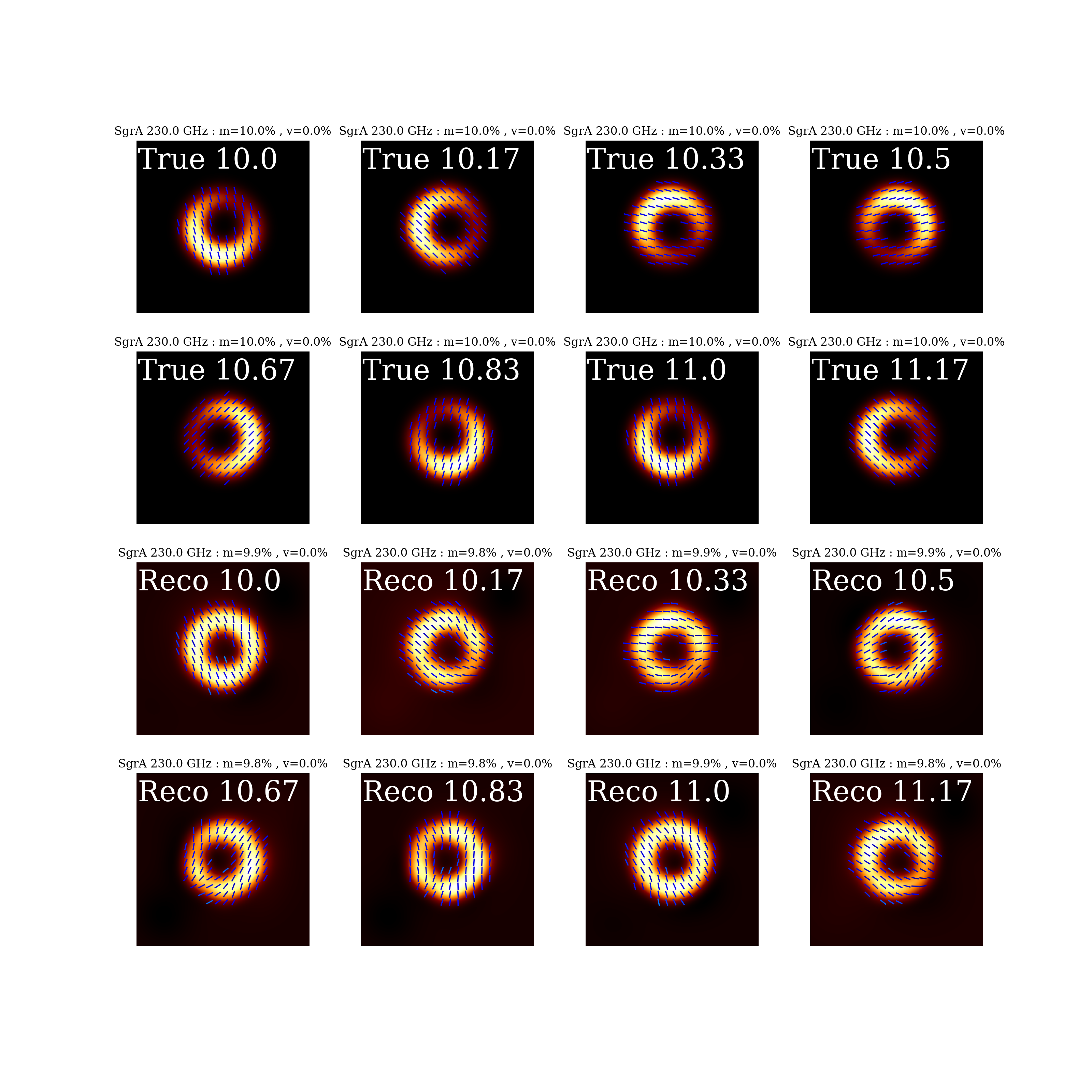}
    \caption{True (upper panels) and recovered (lower panels) test images with full Stokes polarization for the slowly rotating crescent. The mr-support imaging approach succeeds in recovering the true large scale orientation of the EVPA.}
    \label{fig: dynamic_pol}
\end{figure*}

The reconstruction results in the time-window $t \in [10UT, 11UT]$ are presented in Fig. \ref{fig: dynamic_pol} for a slowly rotating crescent model with EHT coverage. The relatively simple polarized structure is well recovered in each frame. While the recovered images show some local variation from the overall orientation, the larger scale EVPA orientation matches for all frames. The fraction of polarized linearly polarized light is surprisingly well recovered. Again, despite of some local variations in the recovered EVPA, the challenging reconstruction does not show image artifacts.

\begin{figure*}
    \centering
    \includegraphics[width=\textwidth]{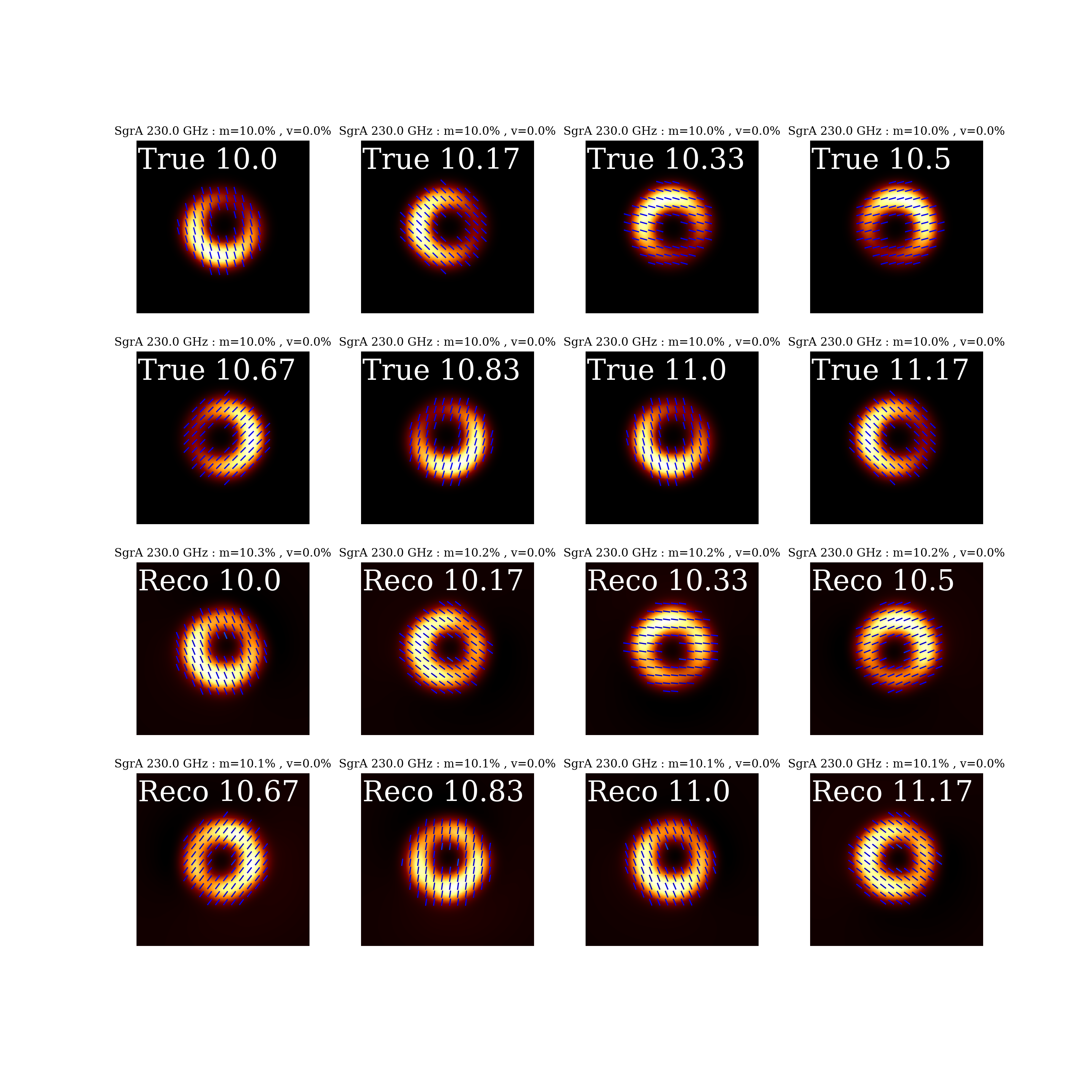}
    \caption{Same as Fig. \ref{fig: dynamic_pol} but with ngEHT coverage: slowly rotating crescent observed with the ngEHT.}
    \label{fig: dynamic_pol_ngeht}
\end{figure*}

In Fig. \ref{fig: dynamic_pol_ngeht} we present the reconstruction of the slowly rotating crescent observed with the ngEHT. The quality of the reconstruction improved compared to the reconstructions presented in Fig. \ref{fig: dynamic_pol}. The global orientation of the EVPAs is well recovered for every frame. In the reconstructions with the EHT configuration we also observed some local variations from the overall polarimetric structure. These cannot be observed anymore in the reconstructions with ngEHT coverage.

\begin{figure*}
    \includegraphics[width=1.3\textwidth,center]{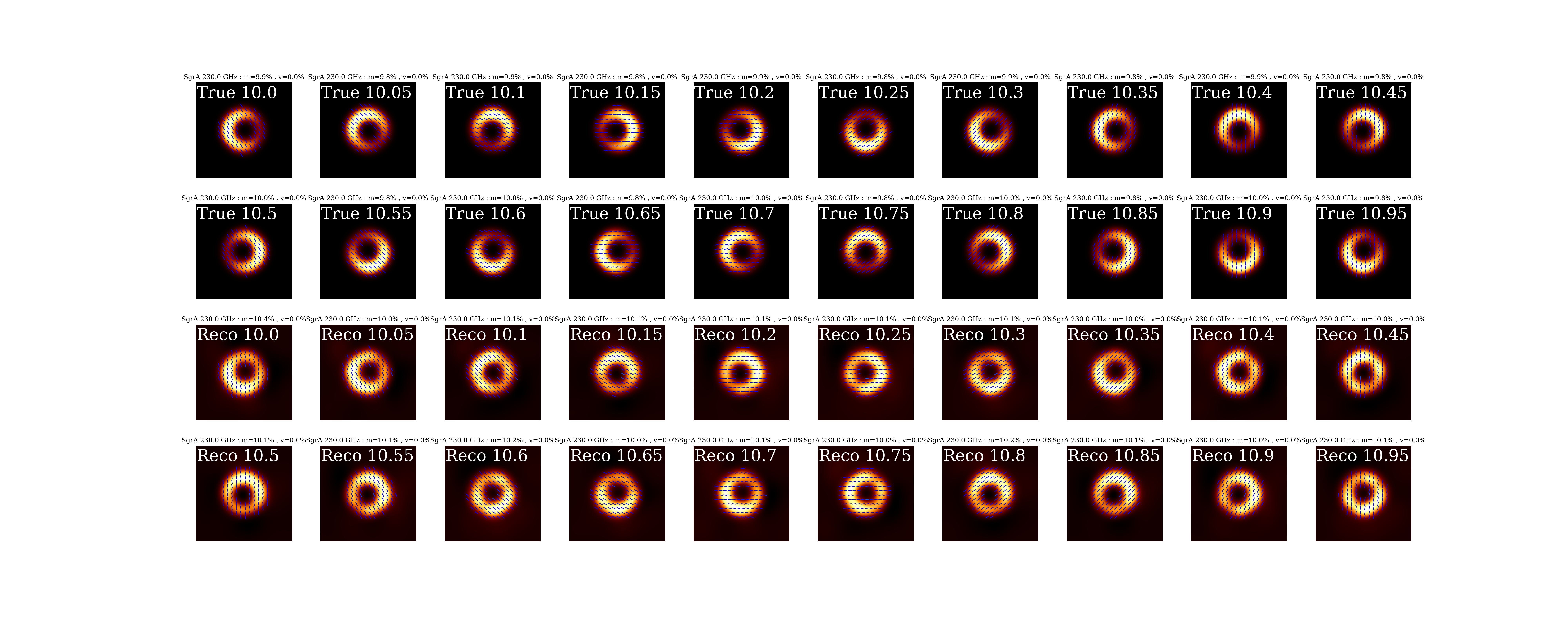}
    \caption{Polarimetric reconstruction of fast rotating crescent with ngEHT coverage.}
    \label{fig: dynamic_pol_ngeht_10_2}
\end{figure*}

We present the dynamic polarimetry reconstruction with mr-support imaging of the rapidly rotating crescent in Fig. \ref{fig: dynamic_pol_ngeht_10_2}. The reconstruction of the polarimetric structure, i.e. the rotation of the EVPAs, remains excellent. These results suggest that mr-support imaging could handle dynamic, polarimetric structural features at the event horizon with realistic dynamic time scales.

\subsection{ngEHT analysis challenge}\label{sec: challenge}
Additionally to the rather simple synthetic data tests presented in the previous subsections, we show here the reconstructions by mr-support imaging for the third ngEHT Analysis challenge \footnote{\url{https://challenge.ngeht.org/}}. The ngEHT Analysis challenges are a series of semi-blind data challenges to evaluate the performance of algorithms for the planned ngEHT instrument \citep{ngehtchallenge}. The ngEHT is a planned instrument to recover (polarimetric) movies at the event horizon scales \citep{Doeleman2019}.

The ground truth movies produced for the ngEHT Analysis challenge resemble the current theoretical state-of-the-art in simulations \citep{ngehtchallenge, Chatterjee2023}. Here we present the reconstructions of a RIAF model of Sgr A* \citep{Broderick2016} with a shearing hotspot \citep{Tiede2020b} with hotspot parameters inspired by \citet{Gravity2018}. The data sets were observed with the EHT array and ngEHT arrays that we used for the geometric data sets as well. In contrast to the proof of concept with geometric models, the ngEHT challenge data contain the full set of data corruptions that may be expected from real observations \citep{ngehtchallenge} simulated with the SYMBA package \citep{Roelofs2020} including atmospheric turbulence, atmospheric opacity, pointing offsets, a scattering screen and thermal noise specific to each antenna. However, no polarization leakage was added to the data. For more information we refer to \citet{ngehtchallenge} and the challenge website \footnote{\url{https://challenge.ngeht.org/}}. The data sets were network calibrated as it is standard in the EHT data processing \citep{eht2022c}. The ngEHT Analysis challenge is in particular well suited as a verification data set since the challenge was done blindly, neither the source files nor the specific data corruptions were made public to the analysis teams.

\begin{figure*}
\includegraphics[width=1.5\textwidth,center]{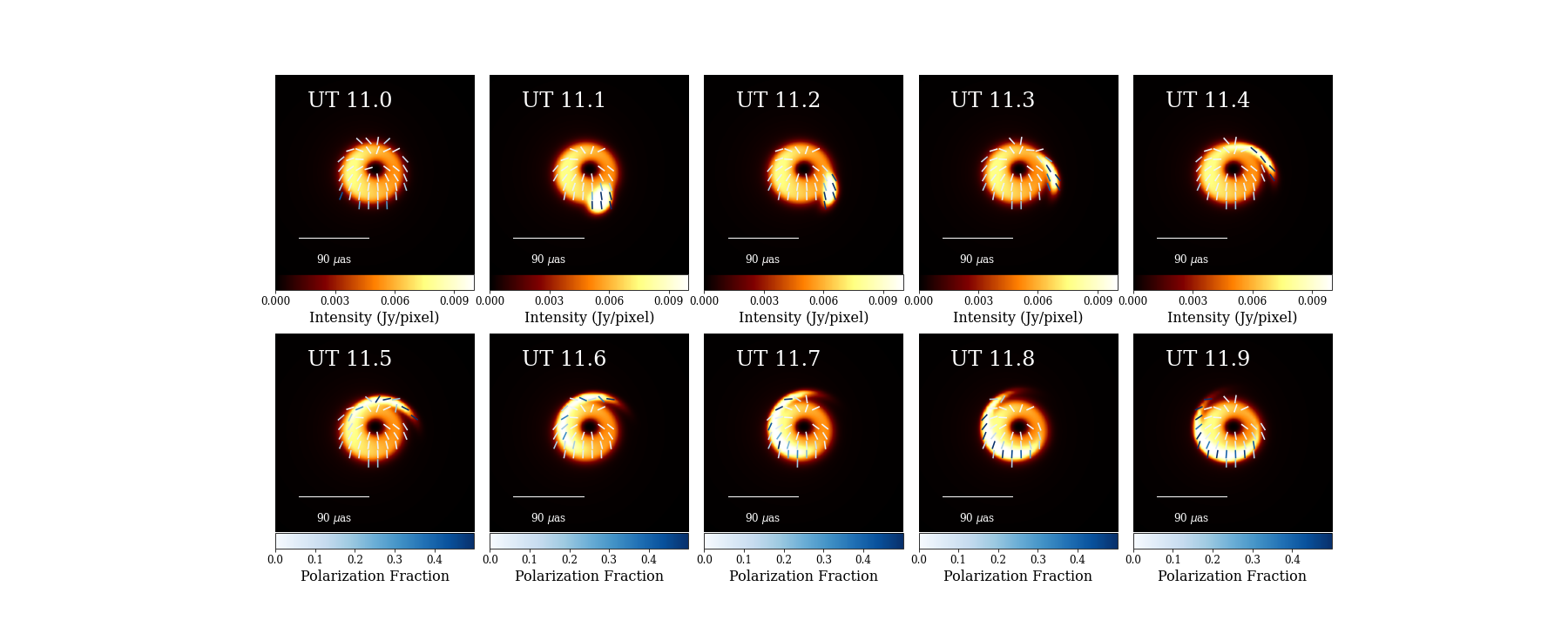}
\caption{Synthetic ground truth movie of Sgr A* used for the third ngEHT Analysis challenge. The model is a RIAF model with a semianalytic shearing hotspot.}
\label{fig: challengegt}
\end{figure*}

We show the ground-truth movie in Fig. \ref{fig: challengegt}. A static (but not descattered) image was recovered by DoG-HiT with a systematic error budget of $2\%$. The static image is computed by DoG-HiT in a completely unsupervised way from closure quantities . We used this calibration independent model to calibrate the data set on long time intervals ($1$ hour). Next we calculated the multiresolution support and cutted the image into frames of six minutes. The dynamic reconstruction was done with mr-support imaging. We self-calibrated the data set in every single observing frame during the procedure. Then we added polarization in every frame.

\begin{figure*}
\includegraphics[width=1.5\textwidth,center]{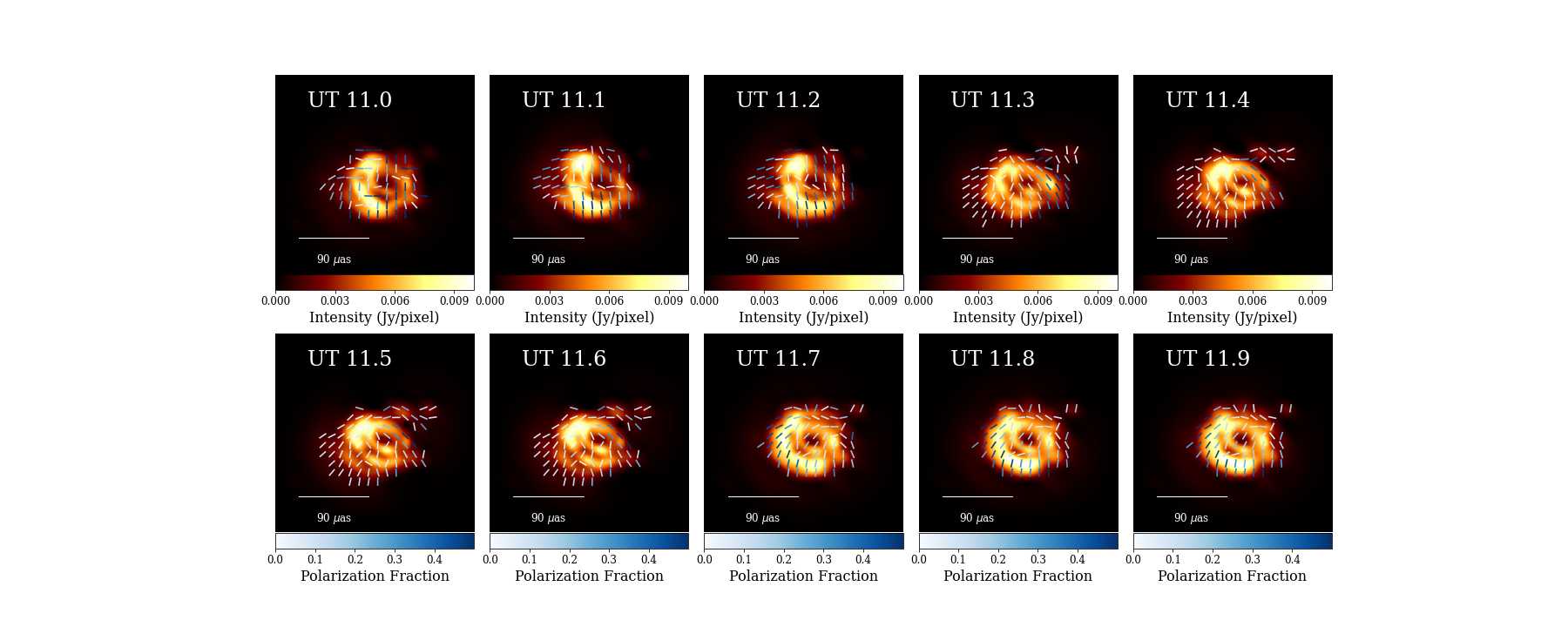}
\caption{Reconstruction of the movie plotted in Fig. \ref{fig: challengegt} with mr-support imaging for the third ngEHT Analysis challenge.}
\label{fig: challengereco}
\end{figure*}

\begin{figure*}
\centering
\includegraphics[width=0.6\textwidth]{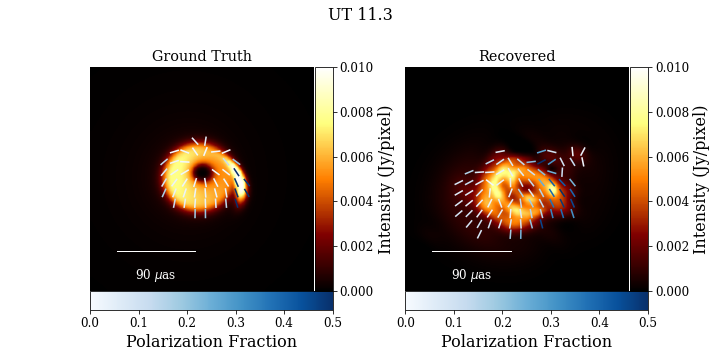} \\
\includegraphics[width=0.6\textwidth]{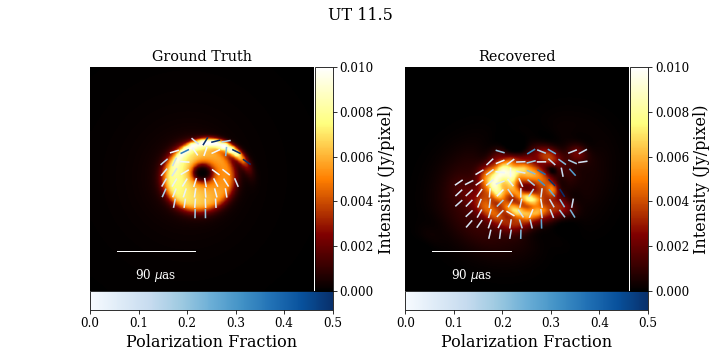} \\
\includegraphics[width=0.6\textwidth]{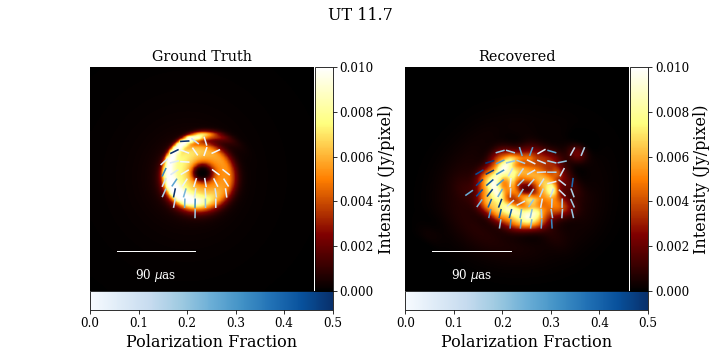}
\caption{Selected frames of the reconstructions shown in Fig. \ref{fig: challengegt} and \ref{fig: challengereco} at times UT 11.3 (upper panels), UT 11.5 (middle panels) and UT 11.7 (lower panels).}
\label{fig: recovsground}
\end{figure*}

The recovered movie is presented in Fig. \ref{fig: challengereco}. Moreover, we show magnified panels of selected frames in Fig. \ref{fig: recovsground}. The single frames all show a ring-like structure with a central depression. Compared to the ground truth frames, the reconstructed images have a worse quality due to the rapid variability, systematics and sparse coverage. Moreover, an interstellar scattering screen was added to the data that was not removed during the imaging procedure. The reconstruction of the shearing hotspot motion is more challenging. We recover an approaching hotspot to the right of the ring at UT 11.3 (upper panels in Fig. \ref{fig: recovsground}), an extended (polarized) tail to the North-West (top right) from UT 11.3 until UT 11.6 (middle panels in Fig. \ref{fig: recovsground}), and a clearly visible arc of larger intensity within the ring to the South-East (bottom left) from UT 11.7-UT 11.9 (bottom panels in Fig. \ref{fig: recovsground}). These features are consistent with the hotspot motion of the ground truth movie. While we recover some motion related to the hotspot motion, a continuously evolving movie was not recovered. This is a result of the rather bad simulated weather conditions and the observation cadence for the third challenge: the source was (synthetically) observed for ten minutes followed by a gap of ten minutes. While mr-support imaging sufficiently recovers some (scattered) hotspot related features in the frames that have observed visibilities, the algorithm does not contain an interpolation scheme to the scans without observations (it just assumes the starting point, i.e. the preceding frame). Hence, we do not recover an evolving movie, but several frames (e.g. UT 11.5 and UT 11.6 or UT 11.7 until end) show the same image structure.

The synthetic ground truth polarization is less dynamic and hence easier to recover. We recover the overall radially-conic EVPA pattern in every frame with minor small scale perturbations from the ground truth (that may be also related to the different Stokes I images). Moreover, the recovered polarization fraction matches the true one. As a more detailed feature we successfully recover a larger fractional polarization for the shearing hotspots that follows the hotspot motion.

The presented data set mimics one of the most challenging VLBI data analysis problems so far with various data corruptions, high frequencies (i.e. phase instabilities), fast dynamics and polarimetric structures, the need for super-resolution, and a sparse VLBI uv-coverage. As expected, the reconstruction quality with mr-support imaging is degraded compared to the rather simple geometric data tests that we discussed before. However, the application highlights already the potential of mr-support imaging to do unsupervised, super-resolving, dynamical and polarimetric imaging together. This presents a unique capability in the landscape of existing imaging algorithms by now, and in particular a domain of research in which CLEAN remains limited due to its lack of resolution, its high demand of human supervision and calibration, and lacking support for dynamical reconstructions.

\section{Conclusions and outlook}
We presented in this manuscript a novel algorithmic approach to do static polarimetry, dynamic imaging and finally dynamic polarimetry. The approach was based on our previous works on multiscalar imaging \citep{Mueller2022, Mueller2022b} and the multiresolution support in particular. The multiresolution support encodes important information about the emission structure on one hand (which spatial scales are present where in the image?) and the uv-coverage on the other hand (which of these spatial scales is measured by baselines?). Hence, the multiresolution support is well suited to introduce regularization for challenging extensions to the standard VLBI imaging problem in the spirit of constrained minimization: we optimize the fit to the respective data terms (chi-squared to frame by frame visibilities or to polarized visibilities), but vary wavelet coefficients in the multiresolution support only.

We demonstrated with applications to simple geometric synthetic observations the power of this approach. The mr-support constraint suppressed the introduction of image artifacts, hence providing ample regularization. Moreover, the approach is flexible enough to allow for the reconstruction of both dynamically evolving structures and polarimetric structures. Moreover, the blind application to more complex movies of the third ngEHT Analysis challenges demonstrated that the algorithm may also provide reasonable reconstructions with real data corruptions in one of the most challenging VLBI imaging problems, although the quality of the reconstruction is degraded.

Mr-support imaging shares the basic advantage of multiscalar approaches that are fitted to the uv-coverage. The static reconstructions are done with DoG-Hit which is completely data-driven and largely automatic without many hyperparameters \citep{Mueller2022}. The same applies for the extension to dynamics and polarimetric quantities. There are no further, specific regularization terms (with corresponding weights) introduced, rather the reconstruction is regularized again by the data driven multiresolution support which is determined by the uv-coverage and baseline-noise. Hence, mr-support imaging is blind and unbiased as well. However, we recognized an important bottlenecks for the dynamic reconstructions with mr-support imaging: the static average image needs to approximate the true time-averaged image quite well.

An extension to RML approaches to dynamic imaging, i.e. the addition of temporal regularizers, is straight-forward as well. Note that due to the lack of regularization parameters controlling the temporal correlation, mr-support imaging basically calculates images with rich structures from the extreme sparsity of a single scan independently of preceding and proceeding scans. That indicates that the multiresolution support information is a rather strong prior information that, once a reasonable static model is established, allows for the handling of extreme sparsity in the data.

The geometric test observations tested throughout this study are rather simple. First, we neglected circular polarization for the purpose of simplicity. We note that we only added thermal noise to the observations and no phase and amplitude errors. This does not affect the reconstruction of the static Stokes I image (neither for a static source nor for a dynamically evolving source) since DoG-HiT uses the closure quantities as data terms only \citep{Mueller2022}. However, phase and amplitude calibration errors could affect the subsequent mr-support imaging rounds since for every frame the (polarized) visibilities are used instead of the closure quantities. Hence, we assume that one was able to solve for the (polarized) self-calibration with the time-averaged mean image. This does not has to be necessarily true, but might be a good approximation when the dynamic time-scale of the source and the dynamic time-scale of the gain-variability are different allowing a gain self-calibration with the mean image \citep[e.g. compare][]{Wielgus2022, eht2022c}.

Moreover, while a rotating crescent movie might be a good approximation to a rotating hotspot model in first instance, the model is only a rough approximation to the range of models for the dynamics at the horizon scale. The same applies to the rather simple polarization model used. We therefore tested the algorithm in the blind third ngEHT Analysis challenge as well. While due to the systematic errors added to the synthetic data, the reconstructions are worse than in the previous data tests, mr-support imaging, for the first time, is able to recover super-resolved, polarized movies in an unsupervised way. This is a unique capability among all currently existing VLBI imaging algorithms. Furthermore, we expect further significant improvements from including a temporal regularizer in the dynamic imaging and from more sophisticated strategies for the static image reconstruction, in particular from frameworks that already demonstrated to be able to recover fast dynamics such as \textit{ehtim} or \textit{StarWarps}.

Finally, the application of the same ground truth movie to a possible ngEHT array configuration demonstrates the improvements that the ngEHT project will bring to dynamic reconstructions. The quality of the fits to Stokes I and polarimetric properties improves. With a ngEHT configuration it is even possible to recover structural patterns with dynamic timescales of about $\sim 10-20\,\mathrm{min}$ and therefore what can be expected from real observations.

\section*{Acknowledgments}
We thank the team from the ngEHT Analysis challenge lead by Freek Roelofs, Lindy Blackburn and Greg Lindahl for the chance to use and publish their synthetic data set for this work. Special thanks goes in particular to Paul Tiede for providing the RIAFSPOT model of SGR A*. HM received financial support for this research from the International Max Planck Research School (IMPRS) for Astronomy and Astrophysics at the Universities of Bonn and Cologne. This work was partially supported by the M2FINDERS project funded by the European Research Council (ERC) under the European Union’s Horizon 2020 Research and Innovation Programme (Grant Agreement No. 101018682). 

Our imaging pipeline and our software is available online as MrBeam software tool\footnote{\url{https://github.com/hmuellergoe/mrbeam}}. Our software makes use of the publicly available ehtim \citep{Chael2018}, regpy \citep{regpy} and WISE software packages \citep{Mertens2015}.

\bibliography{lib}{}
\bibliographystyle{aa}

\appendix

\clearpage

\section{Wavelet dictionaries}
This section is adapted from \citep{Mueller2022b}. The dictionaries used in this paper are as follows::
\begin{align*}
    \Psi^{DoG}: I \mapsto [&G^r_{\sigma_0} * I - G^e_{\sigma_0, \sigma_{1}, \alpha_0} * I, G^r_{\sigma_0} * I - G^e_{\sigma_0, \sigma_{1}, \alpha_1} * I, ..., G^r_{\sigma_0} * I - G^e_{\sigma_0, \sigma_{1}, \alpha_{N-1}} * I, \sum_{i=0}^{N-1}  G^e_{\sigma_0, \sigma_{1}, \alpha_{i}} * I - G^r_{\sigma_1} * I, \\
    &G^r_{\sigma_1} * I - G^e_{\sigma_1, \sigma_{2}, \alpha_{0}} * I, \hspace{1.6cm} ... \hspace{1.6cm}, G^r_{\sigma_1} * I - G^e_{\sigma_1, \sigma_{2}, \alpha_{N-1}} * I, \sum_{i=0}^{N-1}  G^e_{\sigma_1, \sigma_{2}, \alpha_{i}} * I - G^r_{\sigma_2} * I,\\
    &G^r_{\sigma_2} * I - G^e_{\sigma_2, \sigma_{3}, \alpha_{0}} * I, \hspace{1.6cm} ... \hspace{1.6cm}, G^r_{\sigma_2} * I - G^e_{\sigma_2, \sigma_{3}, \alpha_{N-1}} * I, \sum_{i=0}^{N-1}  G^e_{\sigma_2, \sigma_{3}, \alpha_{i}} * I - G^r_{\sigma_3} * I,\\
    & \vdots \\
    &G^r_{\sigma_{J-1}} * I - G^e_{\sigma_{J-1}, \sigma_{J}, \alpha_{0}} * I, \hspace{1.6cm} ... \hspace{1.6cm}, G^r_{\sigma_{J-1}} * I - G^e_{\sigma_{J-1}, \sigma_{J}, \alpha_{N-1}} * I, \sum_{i=0}^{N-1}  G^e_{\sigma_{J-1}, \sigma_{J}, \alpha_{i}} * I - G^r_{\sigma_J} * I,\\
    &G^r_{\sigma_J} * I],
\end{align*}
where $G^r_\sigma$ denotes a radial Gaussian function with a standard deviation $\sigma$ and $G^e_{\sigma_1, \sigma_2, \alpha}$ an elliptical Gaussian with major semiaxis $\sigma_1$, minor semiaxis $\sigma_2$ and angle $\alpha$. The DoB dictionary is composed in the same way by replacing Gaussians with spherical Bessel functions.

\begin{align*}
    \Psi^{DoB}: I \mapsto [&\tilde{J}^r_{\tilde{\sigma}_0} * I - G^e_{\tilde{\sigma}_0, \tilde{\sigma}_{1}, \alpha_0} * I, \tilde{J}^r_{\tilde{\sigma}_0} * I - \tilde{J}^e_{\tilde{\sigma}_0, \tilde{\sigma}_{1}, \alpha_1} * I, ..., \tilde{J}^r_{\tilde{\sigma}_0} * I - \tilde{J}^e_{\tilde{\sigma}_0, \tilde{\sigma}_{1}, \alpha_{N-1}} * I,  \sum_{i=0}^{N-1}  \tilde{J}^e_{\tilde{\sigma}_0, \tilde{\sigma}_{1}, \alpha_{i}} * I - \tilde{J}^r_{\tilde{\sigma}_1} * I, \\
    &\tilde{J}^r_{\tilde{\sigma}_1} * I - \tilde{J}^e_{\tilde{\sigma}_1, \tilde{\sigma}_{2}, \alpha_{0}} * I, \hspace{1.6cm} ... \hspace{1.6cm}, \tilde{J}^r_{\tilde{\sigma}_1} * I - \tilde{J}^e_{\tilde{\sigma}_1, \tilde{\sigma}_{2}, \alpha_{N-1}} * I, \sum_{i=0}^{N-1}  \tilde{J}^e_{\tilde{\sigma}_1, \tilde{\sigma}_{2}, \alpha_{i}} * I - \tilde{J}^r_{\tilde{\sigma}_2} * I,\\
    &\tilde{J}^r_{\tilde{\sigma}_2} * I - \tilde{J}^e_{\tilde{\sigma}_2, \tilde{\sigma}_{3}, \alpha_{0}} * I, \hspace{1.6cm} ... \hspace{1.6cm}, \tilde{J}^r_{\tilde{\sigma}_2} * I - \tilde{J}^e_{\tilde{\sigma}_2, \tilde{\sigma}_{3}, \alpha_{N-1}} * I, \sum_{i=0}^{N-1}  \tilde{J}^e_{\tilde{\sigma}_2, \tilde{\sigma}_{3}, \alpha_{i}} * I - \tilde{J}^r_{\tilde{\sigma}_3} * I,\\
    & \vdots \\
    &\tilde{J}^r_{\tilde{\sigma}_{J-1}} * I - \tilde{J}^e_{\tilde{\sigma}_{J-1}, \tilde{\sigma}_{J}, \alpha_{0}} * I, \hspace{1.6cm} ... \hspace{1.6cm}, \tilde{J}^r_{\tilde{\sigma}_{J-1}} * I - \tilde{J}^e_{\tilde{\sigma}_{J-1}, \tilde{\sigma}_{J}, \alpha_{N-1}} * I, \sum_{i=0}^{N-1}  \tilde{J}^e_{\tilde{\sigma}_{J-1}, \tilde{\sigma}_{J}, \alpha_{i}} * I - \tilde{J}^r_{\tilde{\sigma}_J} * I,\\
    &\tilde{J}^r_{\tilde{\sigma}_J} * I],
\end{align*}
with radial spherical Bessel function $\tilde{J}^r$ and elliptical Bessel function $\tilde{J}^e$. 

\end{document}